\documentclass[nofootinbib,aps,11pt]{revtex4-1}
\usepackage{graphicx}
\usepackage{hyperref}
\usepackage{cancel}
\usepackage{amssymb}
\usepackage{textcomp}
\usepackage{amsmath}
\usepackage{bm}
\usepackage{times}
\usepackage{epsfig}
\usepackage{color}
\newcommand{\hc}{{\rm h.c.}}
\newcommand{\GeV}{{\rm GeV}}
\newcommand{\MeV}{{\rm MeV}}
\newcommand{\TeV}{{\rm TeV}}
\newcommand{\eV}{{\rm eV}}
\newcommand{\keV}{{\rm keV}}
\newcommand{\fb}{{\rm fb}}

\begin{document}
\title{\LARGE Global $U(1)_{L}$ Breaking in Neutrinophilic 2HDM: From LHC Signatures to X-Ray Line}
\bigskip
\author{Weijian Wang~$^{a}$}
\email{wjnwang96@aliyun.com}
\author{Zhi-Long Han~$^{b}$}
\email{hanzhilong@mail.nankai.edu.cn}

\affiliation{ $^a$~Department of Physics, North China Electric Power
University, Baoding 071003,China
\\
$^b$~School of Physics, Nankai University, Tianjin 300071, China}
\date{\today}

\begin{abstract}
Lepton number violation plays an essential role in many scenarios of
neutrino mass generation and also provides new clues to search new
physics beyond the standard model. We consider the neutrinophilic
two-Higgs-doublet model ($\nu$-2HDM) where additional right-handed neutral
fermions $N_{Ri}$ and a complex singlet scalar $\sigma$ are also involved.
In scalar sector, the global $U(1)_{L}$ symmetry is spontaneous broken, leading to
Nambu-Goldstone boson, the Majoron $J$, accompanied by the Majorana
neutrino mass generation. We find that the massless Majoron will induce
large invisible Higgs decay, and current experiments have already set
constraints on relevant  parameters. For the first time, we point out that the
$\nu$-2HDM with $N_{Ri}$ can be distinguished from other seesaw by the same
sign tri-lepton signature $3\ell^\pm4j+\cancel{E}_T$. More interesting, for
$\mathcal{O}(\keV)$ scale Majoron, it is a good candidate of decaying dark matter to
interpret the $3.5~\keV$ and $511~\keV$ line excesses by two different parameter spaces.
\end{abstract}

\maketitle

\section{Introduction}\label{intro}
In standard model, the total lepton number is conserved at classical
level, yet it is violated in many scenarios beyond the standard
model. A widely discussed scenario of the lepton number violation
(LNV) appears in the models for neutrino mass generation. To explain
the no-zero but tiny neutrino mass, the dimension-5 effective
operator $f(\Phi L)(\Phi L)/\Lambda$\cite{Weinberg:1979sa} is introduced so
that the smallness of neutrino mass is attributed to the seesaw
mechanism\cite{type1,type2,type3} where lepton number is violated at a
scale higher than electroweak scale.

The mechanism of LNV may play a key role in the dark side of our
universe. The point is that the pseudo-Nambu-Goldstone boson(pNGB),
the Majoron $J$, arises from the spontaneous breaking of global
$U(1)_{L}$ symmetry\cite{majoron} and picks light mass from quantum
gravitational effect\cite{Coleman:1988tj,Kallosh:1995hi}. In
Ref.\cite{Bazzocchi:2008fh, Esteves:2010sh}, Majoron as a keV dark
matter (DM) candidate has been studied where high LNV scale
(typically $10^{3}-10^{6}$TeV) is required to guarantee the small
coupling of Majoron with neutrinos and eventually produce a
satisfactory DM relic density. Moveover, at one loop level there
exists a sub-leading decay of the Majoron to two photons from its
coupling to charged fermions, leading to further constraints from x-
and $\gamma$-ray experiments. On the other hand, for a TeV LNV
scale, the coupling of Majorons to standard model Higgs boson could
be large. As a result, the new invisible decay modes of Higgs boson
to Majorons is open and provide an interesting route to probe new
physics at LHC\cite{Bonilla:2015uwa,Bonilla:2015jdf}. The
possibility of Majoron as WIMP DM has also been studied in
Ref.\cite{Gu:2010ys,Queiroz:2014yna} where a soft $U(1)_{L}$
breaking term is added to generate the Majoron mass.

In this paper, we investigate the LNV effect in the context of
neutrinophilic two-Higgs-doublet model ($\nu$-2HDM)
\cite{Ma:2000cc,Gabriel:2006ns,Davidson:2009ha,Haba:2011nb} where
one scalar doublet $\Phi$ gives masses to standard model fermions,
while the other scalar doublet $\Phi_{\nu}$ with small vacuum
expectation value (VEV) generates the Dirac neutrino mass term. In
fermion sector, the neutral right-handed fermion singlets $N_{Ri}$
are introduced to give a natural suppression for the light Majorana
neutrino masses. Different from the conventional type-I seesaw
model\cite{type1}, lepton numbers of $N_{Ri}$ are set to be zero
instead of one. In scalar sector, in addition to the SM doublet
scalar $\Phi$, a doublet scalar $\Phi_{\nu}$ with lepton number
$L=1$ and a singlet scalar $\sigma$ with $L=1/2$ are also required
to produce the spontaneous LNV process. Hence the scheme we proposed
can be called ``122" seesaw model in comparison with the ``123"
seesaw model proposed in Ref.\cite{Bazzocchi:2008fh, Esteves:2010sh}
where the ``3" denotes the triplet scalar $\Delta$ in type-II seesaw
\cite{type2}.

The scale of LNV is still unknown, hence both low scale and high
scale scenarios are considered in this work. In former case, the new
massive particles are naturally with electroweak (EW) scale, and
thus contribute rich phenomenon at LHC. For instance, a distinct
same sign trilepton $3\ell^\pm4j+\cancel{E}_T$  signature arising
from the associated production of neutrinophilic scalars is unique,
and therefore making this model quite distinguishable. While for the
massless Majoron, it will contribute to invisible decays of Higgs.
By choosing certain parameters, we find that a large branching ratio
of invisible Higgs decay is possible to escape current experimental
constraints. In the scenario with high LNV scale, we postulate the
existence of $\mathcal{O}(\keV)$-$\mathcal{O}(\MeV)$ Majoron
particle, which serves as a late-decaying dark matter. We find that
the Majoron can decay into two photons. Hence the current
experimental results of X-ray background can set the emission line
constraints on the relevant parameters. As already pointed out in
Ref.\cite{Queiroz:2014yna}, the 3.5keV x-ray line observed by
XMM-Newton observatory\cite{Bulbul:2014sua} can be naturally
explained by $J\to \gamma\gamma$. In addition, we further consider
the 511 keV line from the galactic bulge observed by INTEGRAL
experiment\cite{Knodlseder:2003sv}. It is suggested that the 511 keV
line can be originated from the annihilation of
positronium\cite{Hooper:2004qf,Picciotto:2004rp,Khalil:2008kp} or
radiative decaying of degenerate fermionic DM\cite{Nomura:2016vxr}.
In our model, we suggest that the 511 keV emission line can be
originated from the decay of Majoron into low energy
electron-positron pairs $J\to e^+e^-$. Then the positrons dissipate
their kinetic energy by collisions with baryon galactic gas and
eventually form the positronium with electrons in the cosmic
dust.\cite{Boehm:2003bt}.

The paper is organized as follows. In Sec.\ref{model}, we introduce the
model and describe the details of the symmetry breaking. Possible
constraints from astrophysics, lepton flavor violation, and direct
collider searches are considered in Sec.\ref{constraint}.
In Sec.\ref{HID}, we discuss the contribution of massless Majoron to
invisible decays of Higgs. Collider signatures, especially the LNV
signatures, are carried out in Sec.\ref{coll}.
 In Sec.\ref{MDM}, we consider the Majoron as decaying dark matter and
X-ray sources, where the $3.5\keV$ and $511\keV$ line excesses are also
interpreted. The conclusions are summarised in Sec.\ref{ccl}.

\section{The 122 Majoron Model}\label{model}

\subsection{The Model}
In addition to SM particles, we introduce a singlet scalar $\sigma$,
a neutrinophilic doublet scalar $\Phi_\nu$, and neutral right-handed
fermions $N_{Ri}$. The representations of new particles are listed
in Table. \ref{particle}, where the fields transform under not only
SM gauge group but also global $U(1)_{L}$ group. The lepton number
assignment in Table. \ref{particle} forbids the interaction
$\overline{L}\widetilde{\Phi}N_{R}$, so that only $\Phi_{\nu}$
couples with $N_{R}$. The quark and charged lepton sector, on the
other hand, are the same as the ones in SM. Thus the FCNCs do not
appear at tree level. The relevant interactions are
\begin{equation}
\mathcal{L}_N=-y\overline{L} \tilde{\Phi}_\nu N_{R} +\frac{1}{2}
\overline{N_{R}^c}m_{N} N_{R} + \hc.
\end{equation}
Without loss of generality, we take the diagonal basis for charged
leptons and $N_{R}$.

\begin{table}[!htbp]\large
\begin{tabular}{c c c c c c}
\hline\hline
\quad Field \quad &\quad Spin\quad &\quad $SU(3)_{c}$ \quad
 &\quad  $SU(2)_{L}$ \quad  & \quad  $U(1)_{Y}$ \quad  & \quad  $U(1)_{L}$ \quad  \\
\hline
$N_{Ri}$ & 1/2& 1& 1& 0& 0\\
$\Phi_{\nu}$ & 0& 1& 2& 1/2& 1\\
$\sigma$ &  0& 1& 1& 0& 1/2\\
\hline\hline
\end{tabular}
\caption{New particles content under $G_{SM}\otimes U(1)_{L}$}
\label{particle}
\end{table}

The complete scalar potential is given by
\begin{eqnarray}\label{vphi}\nonumber
V &=& - \mu_2^2 \Phi^\dag\Phi+ \mu_3^2 \Phi^\dag_\nu \Phi_\nu
    +\lambda_1 (\Phi^\dag\Phi)^2 + \lambda_2 (\Phi^\dag_\nu \Phi_\nu)^2
    + \lambda_3 (\Phi^\dag\Phi)(\Phi^\dag_\nu \Phi_\nu)+ \lambda_4 (\Phi^\dag\Phi_\nu)(\Phi^\dag_\nu \Phi)\\
 && - \mu_1^2 \sigma^\dag\sigma + \beta_1 (\sigma^\dag\sigma)^2 + \beta_2 (\Phi^\dag\Phi)(\sigma^\dag\sigma) + \beta_3 (\Phi^\dag_\nu\Phi_\nu)(\sigma^\dag\sigma)
   - k (\Phi^\dag\Phi_\nu\sigma^2 + \hc).
\end{eqnarray}
where after acquiring non-zero VEVs, the scalars are denoted as
\begin{align}
\sigma=\frac{v_{1}+R_{1}+iI_{1}}{\sqrt{2}},~ \Phi=\left(
\begin{array}{c}
\phi^+\\
\frac{v_{2}+R_{2}+iI_{2}}{\sqrt{2}}
\end{array}\right),~
\Phi_\nu=\left(
\begin{array}{c}
\phi^+_\nu\\
\frac{v_{3}+R_{3}+iI_{3}}{\sqrt{2}}
\end{array}\right).
\end{align}
The VEV of $\sigma$ breaks the global symmetry $U(1)_L$
spontaneously through the last term in Eq. \ref{vphi} and also
accounts for the generation of Majorana neutrino masses. The
minimization conditions are given by
\begin{eqnarray}\nonumber\label{mic}
\mu_{1}^{2}&=&\frac{-2\beta_{1}v_{1}^{3}-\beta_{2}v_{2}^{2}v_{1}
   -\beta_{3}v_{3}^{2}v_{1}+2kv_{1}v_{2}v_{3}}{2v_{1}}\\
\mu_{2}^{2}&=&\frac{-2\lambda_{1}v_{2}^{3}-\lambda_{3}v_{3}^{2}v_{2}
   -\lambda_{4}v_{3}^{2}v_{2}-\beta_{2}v_{1}^{2}v_{3}+kv_{1}^{2}v_{3}}{2v_{2}}\\\nonumber
\mu_{3}^{2}&=&\frac{-2\lambda_{2}v_{3}^{3}-\lambda_{3}v_{2}^{2}v_{3}
   -\lambda_{4}v_{2}^{2}v_{3}-\beta_{3}v_{1}^{2}v_{3}+kv_{1}^{2}v_{2}}{2v_{3}}
\end{eqnarray}
Taking the parameter set $\mu_{1,2,3}^{2}>0$ and $k\ll 1$, one can
derive the VEV of $\Phi_{\nu}$ from Eq.\eqref{mic} as following
\begin{equation}
v_{3}\simeq \frac{kv_{1}^{2}v_{2}}{2\mu_{3}^{2}+\beta_{3}v_{1}^{2}}.
\end{equation}
One notes that for $\mu_{3}\ll v_{1}$, we have
\begin{equation}
v_{3}\simeq \frac{k}{\beta_{3}}v_{2},
\end{equation}
where $v_{3}$ is independent to the LNV scale $v_{1}$. For
$\mu_{3}\gg v_{1}$, we have
\begin{equation}
v_{3}\simeq k\frac{v_{1}^{2}}{\mu_{3}^{2}}v_{2}.
\end{equation}
Since $v_3$ is tightly related to tiny neutrino masses, then one
expects the VEV hierarchy $v_{3}\ll v_{2}$ in the condition of
smallness of $k$ or $\mu_{3}\gg v_{1}$. Notably,
$k\Phi^\dag\Phi_\nu\sigma^2$ term is the only source of $U(1)_L$
breaking, radiative corrections to $k$ are proportional to $k$
itself and are only logarithmically sensitive to the cutoff
\cite{Davidson:2009ha}. Thus, the VEV hierarchy $v_3\ll v_2\lesssim
v_1$ is stable against radiative corrections \cite{Haba:2011fn}.

\subsection{Neutrino Masses and Mixing}

\begin{figure}[!htbp]
\begin{center}
\includegraphics[width=0.45\linewidth]{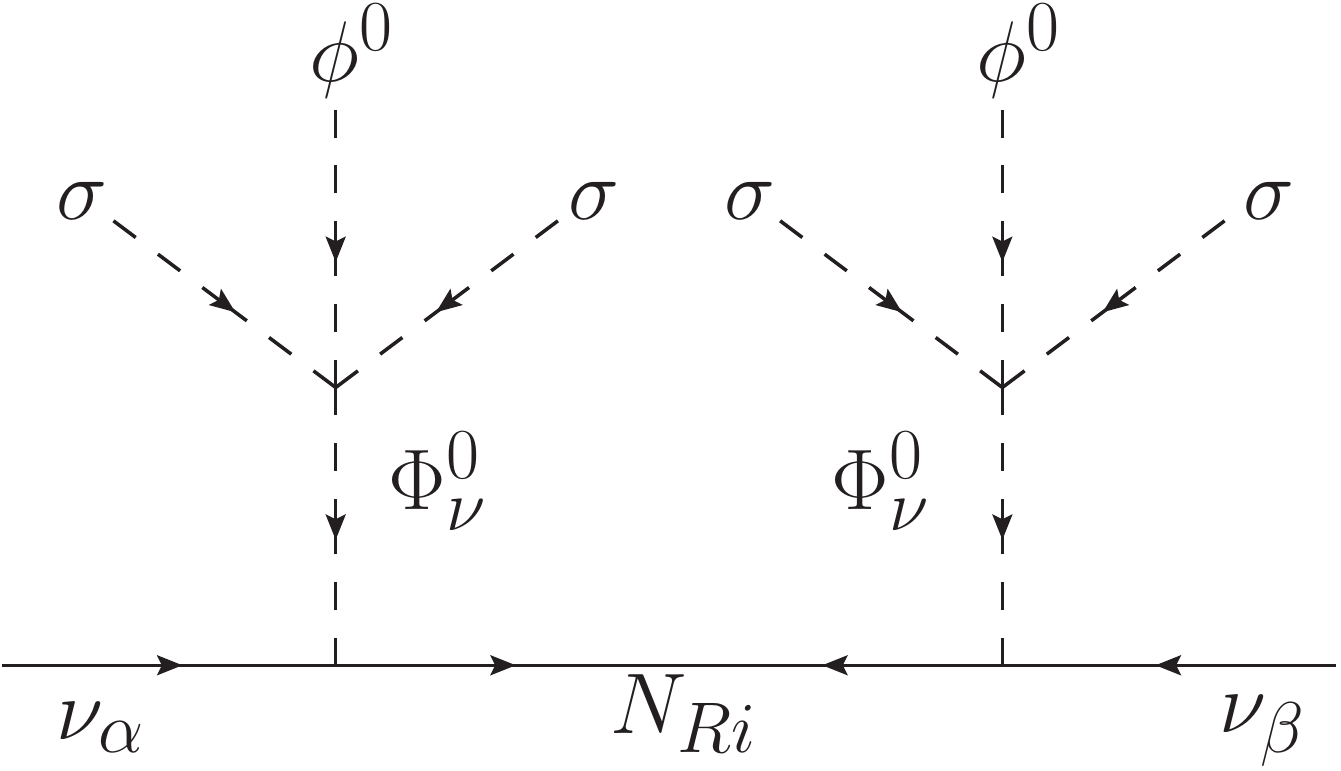}
\end{center}
\caption{The diagram for tree-level neutrino masses in our model.
\label{fig:mv}}
\end{figure}

Note that the term as $\lambda_5/2[(\Phi^\dag\Phi_\nu)^2+\hc]$,
which is allowed by discrete $\mathbb{Z}_2$ symmetry, is now
forbidden by the global $U(1)_L$ symmetry in our model. As shown in
Refs. \cite{Haba:2011nb,Ma:2006km}, such $\lambda_5$ term will
contribute to one-loop induced neutrino masses, and the radiative
induced neutrino masses would be dominant when $\lambda_5
v_2^2/(4\pi)^2\gtrsim v_3^2$. Due to the forbiddance of such
$\lambda_5$ term in our model, the neutrino masses are totally
dominantly induced at tree-level as depicted in Fig. \ref{fig:mv}.
Analogical to canonical Type-I seesaw \cite{type1},  the mass matrix
for light neutrinos can be written as
\begin{equation}\label{eq:mv}
m_\nu = - \frac{v_3^2}{2} y~ m_N^{-1} y^T = U_{\mbox{PMNS}}\, \hat{m}_\nu U^T_{\mbox{PMNS}},
\end{equation}
where $\hat{m}_\nu=\mbox{diag}(m_1,m_2,m_3)$ is the diagonalized neutrino mass matrix, and $U_{\mbox{PMNS}}$ is the PMNS (Pontecorvo-Maki-Nakagawa-Sakata) matrix:
\begin{align}
U_{\mbox{PMNS}} = \left(
\begin{array}{ccc}
c_{12} c_{13} & s_{12} c_{13} & s_{13} e^{i\delta}\\
-s_{12}c_{23}-c_{12}s_{23}s_{13}e^{-i\delta} & c_{12}c_{23}-s_{12}s_{23}s_{13} e^{-i\delta} & s_{23}c_{13}\\
s_{12}s_{23}-c_{12}c_{23}s_{13}e^{-i\delta} & -c_{12}s_{23}-s_{12}c_{23}s_{13}e^{-i\delta} & c_{23}c_{13}
\end{array}
\right)\times
\left(
\begin{array}{ccc}
e^{i\varphi_1/2} & 0 &0 \\
0 & e^{i\varphi_2/2} & 0 \\
0 & 0 & 1
\end{array}
\right)
\end{align}
Here, we use $c_{ij}=\cos\theta_{ij}$ and $s_{ij}=\sin\theta_{ij}$ for short, $\delta$ is the Dirac phase and $\varphi_1,\varphi_2$ are the two Majorana phases. In the following numerical discussion of the phenomenology, we take into account both normal (NH) and inverted hierarchy (IH), and use the latest best fit values of neutrino oscillation parameters in Ref. \cite{Gonzalez-Garcia:2014bfa} \footnote{Early works on the global fit of neutrino oscillation can be found in Refs. \cite{Tortola:2012te}.}. For simplicity, the Majorana phases $\varphi_{1,2}$ are neglected in the following numerical discussion. According to Eq. \ref{eq:mv}, the Yukawa matrix $y$ can be expressed in terms of quantities measured in neutrino oscillation experiments. Since the neutrino masses are induced by Type-I seesaw like mechanism in our model, we could adopt the Casas-Ibarra parametrization \cite{Casas:2001sr} to express $y$ as:
\begin{equation}
y=\frac{\sqrt{2}}{v_3}U_{\mbox{PMNS}}\sqrt{\hat{m}_\nu} R \sqrt{m_N},
\end{equation}
where $R$ is a complex orthogonal matrix. In the minimal case for two massive neutrinos, $R$ can be expressed in terms of an angle $\omega$ \cite{Ibarra:2003up} as:
\begin{align}
R^{\mbox{NH}}=\left(
\begin{array}{cc}
0 & 0 \\
\sqrt{1-\omega^2} & -\omega \\
\omega & \sqrt{1-\omega^2}
\end{array}
\right),\quad
R^{\mbox{IH}}=\left(
\begin{array}{cc}
\sqrt{1-\omega^2} & -\omega \\
\omega & \sqrt{1-\omega^2} \\
0 & 0
\end{array}
\right),
\end{align}
for the normal (NH) and inverted (IH) hierarchy, respectively. Here
in this work, we concentrate on the range $-1<\omega<1$. Typically,
for $v_3\sim1\MeV$, $m_\nu\sim0.1~\eV$ and $m_{N}\sim100~\GeV$, we
have $y\sim0.01$.

\subsection{Scalar Masses and Mixings}
The squared mass matrix for neutral CP-even scalars in the weak
basis $(R_{1},R_{2},R_{3})$ is below
\begin{equation} M_{R}^{2}=\left(\begin{array}{ccc}
2\beta_{1}v_{1}^{2}&\beta_{2}v_{1}v_{2}-kv_{1}v_{3}&\beta_{3}v_{1}v_{3}-kv_{1}v_{2}\\
\beta_{2}v_{1}v_{2}-kv_{1}v_{3}&2\lambda_{1}v_{2}^{2}+\frac{1}{2}kv_{1}^{2}\frac{v_{3}}{v_{2}}&(\lambda_{3}+\lambda_{4})v_{2}v_{3}-\frac{1}{2}kv_{1}^{2}\\
\beta_{3}v_{1}v_{3}-kv_{1}v_{2}&(\lambda_{
3}+\lambda_{4})v_{2}v_{3}-\frac{1}{2}kv_{1}^{2}&2\lambda_{2}v_{3}^{2}+\frac{1}{2}kv_{1}^{2}\frac{v_{2}}{v_{3}}
\end{array}\right).
\label{nrs}
\end{equation}
The $M_{R}^{2}$ is diagonalized by orthogonal matrix $O^{R}$ as
$O^{R}M_{R}^{2}(O^{R})^{T}=\mbox{diag}(m_{H_{1}}^{2},m_{H_{2}}^{2},m_{H_{3}}^{2})$,
where
\begin{equation}
\left(\begin{array}{c}
H_{1}\\
H_{2}\\
H_{3}
\end{array}\right)=O^{R}\left(\begin{array}{c}
R_{1}\\
R_{2}\\
R_{3}
\end{array}\right),
\end{equation}
and $O^{R}$ is parameterized as
\begin{equation}
O^{R}=\left(\begin{array}{ccc}
c_{12}c_{13}&c_{13}s_{12}&s_{13}\\
-s_{12}c_{23}-c_{12}s_{13}s_{23}&c_{12}c_{23}-s_{12}s_{13}s_{23}&c_{13}s_{23}\\
s_{23}s_{12}-c_{12}c_{23}s_{13}&-c_{12}s_{23}-c_{23}s_{12}s_{13}&c_{13}c_{23}
\end{array}\right),
\label{or}
\end{equation}
with $c_{ij}=\cos\alpha_{ij}$ and $s_{ij}=\sin\alpha_{ij}$ for
short.  In our following discussion, we will always keep
$H_2$ to be the discovered standard model (SM) like Higgs boson with
$m_{H_2}=125\GeV$ at LHC
\cite{Aad:2012tfa,Chatrchyan:2012xdj,Aad:2015zhl}.

The squared mass matrix for CP-odd scalars in the basis of
$(I_{1},I_{2},I_{3})$ is given by
\begin{equation} M_{I}^{2}=k\left(\begin{array}{ccc}
2v_{2}v_{3}&-v_{1}v_{3}&v_{1}v_{2}\\
-v_{1}v_{3}&\frac{1}{2}v_{1}^{2}\frac{v_{3}}{v_{2}}&-\frac{1}{2}v_{1}^{2}\\
v_{1}v_{2}&-\frac{1}{2}v_{1}^{2}&\frac{1}{2}v_{1}^{2}\frac{v_{2}}{v_{3}}
\end{array}\right)
\end{equation}
The matrix $M_{I}^{2}$ is diagonalized as
$O^{I}M_{I}^{2}(O^{I})^{T}=\mbox{diag}(0,0,m_{A}^{2})$, where
\begin{equation}
\left(\begin{array}{c}
J\\
G^{0}\\
A
\end{array}\right)=O^{I}\left(\begin{array}{c}
I_{1}\\
I_{2}\\
I_{3}
\end{array}\right).
\end{equation}
As one could expect, two eigenstates with null masses are obtained,
corresponding to the normal SM Goldstone boson $G^{0}$ and the
Majoron $J$ generated from global LNV. The $m_{A}^{2}$ and matrix
$O^{I}$ are given by
\begin{equation}
m_{A}^{2}=k\Big(\frac{v_{1}^{2}v_{3}^{2}+4v_{3}^{2}v_{2}^{2}+v_{1}^{2}v_{2}^{2}}{2v_{2}v_{3}}\Big),
\label{as}
\end{equation}
\begin{equation}
O^{I}=\left(\begin{array}{ccc}
cv_{1}V^{2}&32cv_{2}v_{3}^{2}&-32cv_{2}^{2}v_{3}\\
0&-\frac{4v_{2}}{V}&-\frac{4v_{3}}{V} \\
-\frac{2bv_{2}}{v_{1}}&b&-\frac{bv_{2}}{v_{3}}
\end{array}\right),
\label{oi}
\end{equation}
with
\begin{equation}\begin{split}
&V^{2}=16(v_{2}^{2}+v_{3}^{2})\\
&c^{-2}=v_{1}^{2}V^{4}+1024(v_{2}^{2}v_{3}^{4}+v_{2}^{4}v_{3}^{2})\\
&b^{2}=\frac{v_{1}^{2}v_{3}^{2}}{v_{2}^{2}v_{1}^{2}+4v_{2}^{2}v_{3}^{2}+v_{3}^{2}v_{1}^{2}}
\end{split}\end{equation}
Turning to the charged scalars, the associated squared mass matrix
in the basis of $(\phi^{\pm}, \phi_\nu^{\pm})$ is given by
\begin{equation} M_{H^{\pm}}^{2}=\frac{1}{2}\left(\begin{array}{cc}
kv_{1}^{2}\frac{v_{3}}{v_{2}}-\lambda_{4}v_{3}^{2}&\lambda_{4}v_{2}v_{3}-kv_{1}^{2}\\
\lambda_{4}v_{2}v_{3}-kv_{1}^{2}&kv_{1}^{2}\frac{v_{2}}{v_{3}}-\lambda_{4}v_{2}^{2}
\end{array}\right)
\end{equation}
Then we have $O^{\pm}M_{H^{\pm}}^{2}(O^{\pm})^{T}=\mbox{diag}(0,m_{H^{\pm}}^{2})$, where
\begin{equation}
\left(\begin{array}{c}
G^{\pm}\\
H^{\pm}
\end{array}\right)=\left(\begin{array}{cc}
c_{\pm}&s_{\pm}\\
-s_{\pm}&c_{\pm}
\end{array}\right)
\left(\begin{array}{c}
\phi^{\pm}\\
\phi_{\nu}^{\pm}
\end{array}\right)
\end{equation}
with
$c_{\pm}=v_{2}/\sqrt{v_{2}^{2}+v_{3}^{2}}$, $s_{\pm}=v_{3}/\sqrt{v_{2}^{2}+v_{3}^{2}}$
and the mass of $m_{H^{\pm}}$ given by
\begin{equation}
m_{H^{\pm}}^{2}=\frac{1}{2v_{2}v_{3}}(v_{2}^{2}+v_{3}^{2})(kv_{1}^{2}-\lambda_{4}v_{2}v_{3})
\label{cs}\end{equation}  Taking into account the smallness of
$v_{3}$ and $k$ one notices from Eq.\eqref{nrs} that for the neutral
scalars $H_{3}$ and $A$, the following mass relation holds
approximately
\begin{equation}
m_{A}^{2}\simeq
kv_{1}^{2}v_{2}/2v_{3}=\big[M_{R}^{2}\big]_{33}\simeq m_{H_{3}}^{2}
\label{ma}\end{equation} In the same way, from Eq.\eqref{as} and
\eqref{cs} one derives the mass relation
\begin{equation}
m_{A}^{2}-m_{H_{+}}^{2}\approx\frac{\lambda_{4}}{2}v_{2}^{2},
\end{equation}
which implies that the differences between $m_{A}$ and $m_{H^{+}}$
can not be two large under perturbativity condition. For simplicity,
we will assume that masses of neutrinophilic scalars are degenerate,
i.e., $m_{H^+}\!=\!m_{H_3}\!=\!m_{A}\equiv\!m_{\Phi_\nu}$.

\section{Constraints}\label{constraint}

\subsection{Theoretical Constraints}
Using Eqs.\eqref{nrs}, \eqref{as} and \eqref{cs}, we
rewrite all the coupling constants $\lambda_{i}$ and $\beta_{j}$ in
terms of mixing angles $\alpha_{ij}$ and scalar masses
\begin{equation}\begin{split}
&\beta_{1}=\frac{1}{2v_{1}^{2}}[M_{R}^{2}]_{11}\\
&\beta_{2}=\frac{1}{v_{1}v_{2}}[M_{R}^{2}]_{12}+k\frac{v_{3}}{v_{1}}\\
&\beta_{3}=\frac{1}{v_{1}v_{3}}[M_{R}^{2}]_{13}+k\frac{v_{2}}{v_{3}}\\
&\lambda_{1}=\frac{1}{2v_{2}^{2}}[M_{R}^{2}]_{22}-k\frac{v_{1}^{2}v_{3}}{4v_{2}^{3}}\\
&\lambda_{2}=\frac{1}{2v_{3}^{2}}[M_{R}^{2}]_{33}-k\frac{v_{1}^{2}v_{2}}{4v_{3}^{3}}\\
&\lambda_{3}=\frac{1}{v_{2}v_{3}}[M_{R}^{2}]_{23}-\lambda_{4}+k\frac{v_{1}^{2}}{2v_{2}v_{3}}
\end{split}\end{equation}
and
\begin{equation}\begin{split}
&\lambda_{4}=\frac{1}{v_{2}v_{3}}(kv_{1}^{2}-\frac{2v_{2}}{v_{2}+v_{3}}m_{H^{\pm}}^{2})\\
&k=\Big(
\frac{2v_{2}v_{3}}{v_{1}^{2}v_{2}^{2}+v_{3}^{2}v_{2}^{2}+4v_{2}^{2}v_{3}^{2}}\Big)m_{A}^{2}
\end{split}\label{k}\end{equation}
where $[M_{R}^{2}]_{ij}$ denotes the matrix elements of $M_{R}^{2}$.

The scalar potential is bounded from below if the quartic part of
scalar potential is positive in the non-negative basis. In the
following, we take the same procedure in Ref.\cite{Bazzocchi:2008fh,
Esteves:2010sh,Bonilla:2015jdf}. Taking into account the fact of
$v_{3}\ll v_{1}$ and using Eq.\eqref{k}, we derive the parameter $k$
as
\begin{equation}
k\approx m_{A}^{2}\frac{2v_{3}}{v_{1}^{2}v_{2}}
\end{equation}
Therefore, we have $k\ll \lambda_{i},\beta_{i}$ and the parameter
$k$ can be neglected with respect to other coupling constants. In
this limit, the $\emph{copositive}$ criteria\cite{kan2012} can be
applied to the quartic part of scalar potential to give the
boundedness condition as following
\begin{equation}\begin{split}\label{bd}
\lambda_{1}>0,\quad\quad\quad \lambda_{2}>0,\quad\quad\quad
\beta_{1}>0\\
x=\lambda_{3}+\theta(-\lambda_{4})\lambda_{4}+2\sqrt{\lambda_{1}\lambda_{2}}>0\\
y=\beta_{2}+2\sqrt{\lambda_{1}\beta_{1}},\quad\quad\quad
z=\beta_{3}+2\sqrt{\lambda_{2}\beta_{1}}\\
\sqrt{\lambda_{1}\lambda_{2}\beta_{1}}+[\lambda_{3}+\theta(-\lambda_{4})\lambda_{4}]\sqrt{\beta_{1}}+\beta_{2}\sqrt{\lambda_{2}}+\beta_{3}\sqrt{\lambda_{1}}+\sqrt{xyz}>0
\end{split}\end{equation}
In additional, we set the values of coupling constants $\lambda_{i}$
and $\beta_{j}$ less then $\sqrt{4\pi}$ to ensure the perturbative
condition.

\subsection{Astrophysical Constraints}

\begin{figure}[!htbp]
\begin{center}
\includegraphics[width=0.45\linewidth]{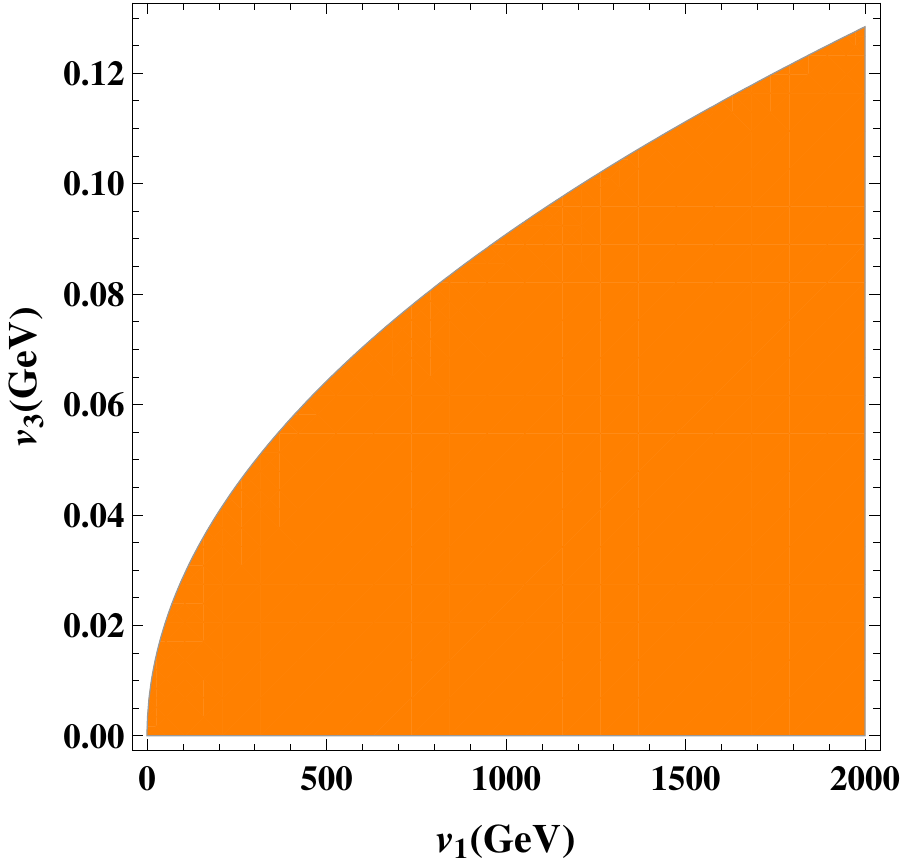}
\end{center}
\caption{Allowed region of $v_3$ as a function of $v_1$ considering the constraint from Majoron-electron coupling.
\label{v3-v1}}
\end{figure}

Note that $\rho=1$ at tree level in this $\nu$2HDM. The stringent
constraint on $v_3$ comes from astrophysics, due to the
contributions of Majoron-electron coupling $g_{Jee}$ to supernova
\cite{Dearborn:1985gp} and red giant cooling \cite{Viaux:2013lha}.
For a massless Majoron (or lighter than typical stellar
temperatures), the Compton-like process $\gamma+e\to J+e$ sets an
upper bound for the $g_{Jee}$ coupling as
\cite{Dearborn:1985gp,Viaux:2013lha}:
\begin{equation}
|g_{Jee}|=|O^I_{12}\frac{m_e}{v_2}|\lesssim1.4\times10^{-13}.
\end{equation}
Considering the profile of Majoron \cite{Schechter:1981cv} in Eq. \ref{Jpro}, we can translate this as a bound on the projection of the Majoron onto the doublet $\Phi$ as \cite{Diaz:1998zg}:
\begin{equation}\label{gjee}
|\langle J|\Phi\rangle|=\frac{2 v_2 v_3^2}{\sqrt{v_1^2(v_2^2+v_3^2)^2+4v_2^2v_3^4+4v_2^4v_3^2}}
\approx\frac{2 v_3^2}{v_1 v_2}\lesssim 6.7\times10^{-8},
\end{equation}
where in above approximation, we have used the assumption that
$v_3\ll v_1,v_2$. So from Eq. \ref{gjee}, we expect that
$v_3^{\tiny\mbox{Max}}\propto\sqrt{v_1}$. The allowed region of
$v_3$ as a function of $v_1$ is presented in FIG. \ref{v3-v1}. For
instance, $v_3\lesssim0.09\GeV$ must be satisfied when
$v_1=1000\GeV$.

\subsection{Lepton Flavor Violation}
\begin{figure}[!htbp]
\begin{center}
\includegraphics[width=0.45\linewidth]{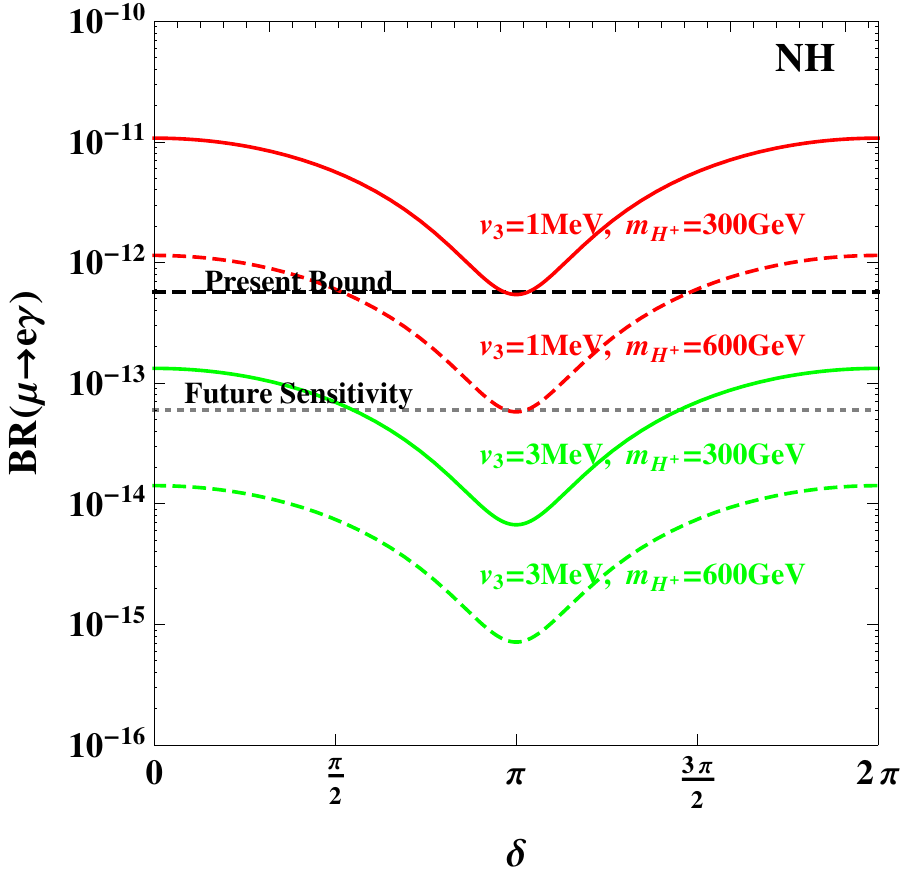}
\includegraphics[width=0.45\linewidth]{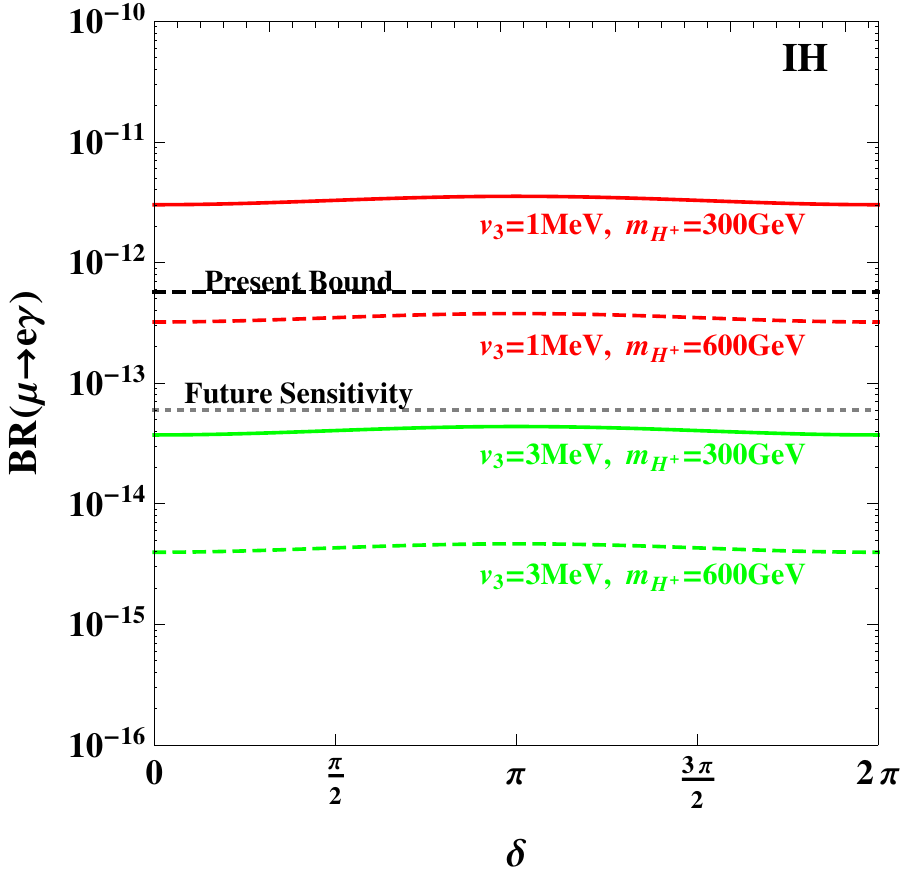}
\end{center}
\caption{BR$(\mu\to e \gamma)$ as a function of $\omega$ for four different values of $v_3$ and $m_{H^+}$ in NH (left panel) and IH (right panel).
\label{LFV}}
\end{figure}

We find that the lepton flavor violating (LFV) processes would set a
much more stringent lower bound on $v_3$ in models with heavy exotic
leptons \cite{Ding:2014nga} than the canonical type-II seesaw
\cite{Fukuyama:2009xk,Han:2015hba} as well as the Dirac neutrino
scenario of $\nu$2HDM \cite{Bertuzzo:2015ada}. In this paper, we
simply take the $\mu\to e\gamma$ process to illustrate such tight
constraints, since the MEG experiment sets a severe upper limit as
BR$(\mu\to e\gamma)<5.7\times10^{-13}$\cite{Adam:2013mnn}. We also
consider the future sensitivity of MEG experiment, which might be
down to $6\times10^{-14}$ \cite{Baldini:2013ke}. The branching ratio
of $\mu \to e \gamma$ is calculated as \cite{Ma:2001mr}:
\begin{equation}\label{BRLFV}
\mbox{BR}(\mu\to e\gamma)=\frac{3\alpha}{64\pi G_F^2}\left|\sum_i \frac{y_{\mu i}y_{e i}^*}{m_{H^+}^2} F\left(\frac{m_{N_i}^2}{m_{H^+}^2}\right)\right|^2,
\end{equation}
where the loop function $F(x)$ is:
\begin{equation}
F(x)=\frac{1-6x+3x^2+2x^3-6x^2\ln x}{6(1-x)^4}.
\end{equation}

In Fig. \ref{LFV}, we show the numerical results of BR($\mu\to e \gamma$) as a function of $\omega$ for ($v_3,m_{H^+}$)=(1MeV,300GeV), (1MeV,600GeV), (3MeV,300GeV) and (3MeV,600GeV) in both normal and inverted hierarchy. In case of normal hierarchy, the present MEG bound \cite{Adam:2013mnn} requires $v_3 m_{H^+}\gtrsim 600\MeV\!\cdot\GeV$, meanwhile the future MEG sensitivity \cite{Baldini:2013ke} would push this bound up to $v_3 m_{H^+}\gtrsim 900\MeV\!\cdot\GeV$. On the other hand in case of inverted hierarchy, the bound on $v_3 m_{H^+}$ is slightly less stringent than the the bound of normal hierarchy. Briefly, we can conclude that to satisfy the LFV constraint, $v_3\gtrsim\mathcal{O}(\MeV)$ is needed for $m_{H^+}\sim\mathcal{O}(\TeV)$.

In general, LFV processes depends on neutrino masses,
mixing angles, Dirac phase, as well as Majorana phases. In our
assumption with degenerate $N_R$ and real $R$ matrix, we obtain
\begin{eqnarray}
\sum_i y_{\mu i}y_{ei}^{*}\propto U_{\text PMNS} \hat{m}_\nu U_{\text PMNS}^\dag
&=& c_{12}c_{13}s_{12}c_{23}(m_{\nu_2}-m_{\nu_1}) \\ \nonumber
&+& c_{13}s_{13}s_{23}e^{-i\delta}[(m_3-m_2)+c_{12}^2(m_2-m_1)]
\end{eqnarray}
Therefore, the $\mu\to e\gamma$ sets no constraint on the Majorana
phases and the $R$ matrix. But for a large $s_{13}$, the branch ratio is sensitive to
the Dirac phase $\delta$.

Comparing to the bound on type-II seesaw $v_{\Delta}m_{H^{++}}\gtrsim 150 \eV\!\cdot\GeV$ \cite{Fukuyama:2009xk} \footnote{$v_{\Delta}$ is the vacuum expectation value of Higgs triplet $\Delta$, and $m_{H^{++}}$ is the mass of the doubly-charged scalar $H^{++}$ in $\Delta$.} and Dirac scenario of $\nu$2HDM $v_3 m_{H^{+}}\gtrsim 250 \eV\!\cdot\GeV$ \cite{Bertuzzo:2015ada}, the bound on the Majorana scenario of $\nu$2HDM $v_3 m_{H^{+}}\gtrsim 600 \MeV\!\cdot\GeV$ is about 6 orders of magnitudes higher. Here we take Dirac and Majorana scenario of $\nu$2HDM to briefly estimate such great difference. From Eq. \ref{BRLFV}, it is clear that the constraint from LFV actually requires about the same order of the Yukawa coupling $y$, since the loop function $F(x)$ is of the same order in both Dirac and Majorana scenario if we also assume $m_{N}<m_{H^+}$. In Dirac scenario, $y^D\sim m_\nu/v_3^D$, while in Majorana scenario, $y^M\sim \sqrt{m_\nu m_N}/v_3^M$. For the same order of the Yukawa coupling, we could estimate that $v_3^M/v_3^D\sim \sqrt{m_N/m_\nu}\sim10^6$ with $m_N\sim10^2\GeV$ and $m_\nu\sim0.1\eV$, which is just the result of the above discussion.

\subsection{Collider Constraints}\label{coco}

The status of the Higgs singlet $H_1$ has been extensively studied in Refs. \cite{Robens:2015gla,Buttazzo:2015bka,Robens:2016xkb,Wang:2015saa}.
We refer to Ref. \cite{Robens:2016xkb} for a more detail and updated study on the constraints of $H_1$. In the high mass region with $m_{H_1}>130\GeV$, the allowed value for $\sin\alpha_{12}$ as a function of $m_{H_1}$ is shown in FIG.1 of Ref. \cite{Robens:2016xkb}. For example, $\sin\alpha_{12}\lesssim0.3$ is required with $m_{H_1}=300\GeV$. Although the invisible decay $H_1\to JJ$ could affect the direct search bound to be less stringent, the indirect bound as from Higgs signal rate still requires $\sin\alpha_{12}<0.36$ \cite{Robens:2016xkb}. So we will consider $\sin\alpha_{12}=0.3,0.2,0.1$ with $m_{H_1}=300\GeV$ as our benchmark points for the high mass region in Sec. \ref{HID}.

\begin{figure}[!htbp]
\begin{center}
\includegraphics[width=0.45\linewidth]{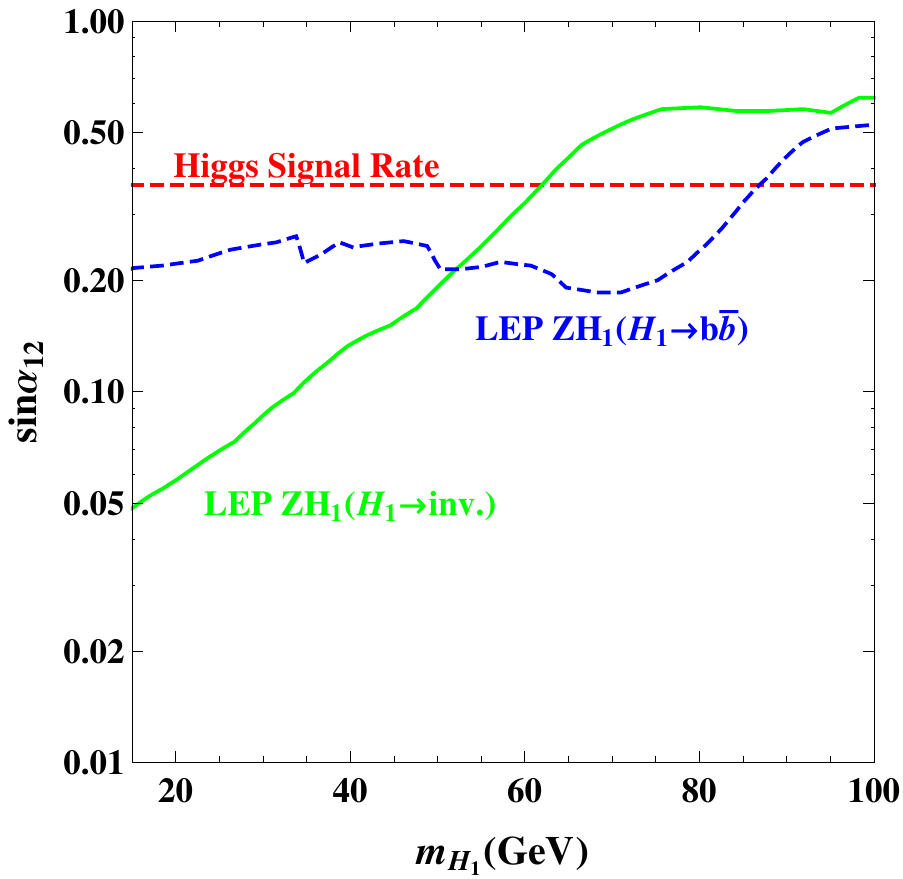}
\end{center}
\caption{Constraints on $\sin\alpha_{12}$ in the low mass region.
\label{Sin12}}
\end{figure}

However, in the low mass region with $m_{H_1}<120\GeV$, the stringent bound in Ref. \cite{Robens:2016xkb} is not applicable to our model. Mainly because the invisible decay $H_1\to JJ$ is totally dominant in this region (see the detail study in Sec. \ref{HID}). In FIG. \ref{Sin12}, we show the constraints on $\sin\alpha_{12}$ in the low mass region coming from LHC Higgs signal rate \cite{Robens:2016xkb}, visible ($H_1\to b\bar{b}$) \cite{Abdallah:2004wy} and invisible decaying ($H_1\to JJ$) \cite{Abbiendi:2007ac} Higgs at LEP through $ZH_{1}$ associated production. Note that for the constraint from $H_1\to b\bar{b}$ we show the most severe case with BR$(H_1\to b\bar{b})=1$. If we take into account a more realistic BR$(H_1\to b\bar{b})$, the exclusion region will be $\sin\alpha_{12}>0.4$ and less stringent than those from Higgs signal rate \cite{Bonilla:2015uwa,Bonilla:2015jdf}.
For $m_{H_1}=50\GeV$, the most strict bound comes from invisible Higgs search at LEP with $\sin\alpha_{12}\lesssim0.2$. Therefore, we will take $\sin\alpha_{12}=0.2,0.1,0.05$ with $m_{H_1}=50\GeV$ as our benchmark points for the low mass region in Sec. \ref{HID}.

The collider signature of $\nu$2HDM has been discussed in Refs. \cite{Haba:2011nb,Davidson:2009ha,Davidson:2010sf}. In the case of $m_N>m_{H^+}$, the dominant decay mode of $H^+$ could be $H^+\to\ell^+\nu$. The direct search for signature as $\ell^+\ell^-+\cancel{E}_T$ at LHC has excluded the region of $m_{H^+}\lesssim300\GeV$ \cite{Aad:2014vma,Khachatryan:2014qwa}. While in the case of $m_N<m_{H^+}$, the dominant decay mode of $H^+$ would be $H^+\to \ell^+ N_{Ri}$ with the heavy Majorana neutrino $N_{Ri}$ further decaying into $\ell^\pm W^\mp$, $\nu Z$ and $\nu H_2$. A detail discussion and simulation at LHC of this case is still missing. Therefor, we consider the LEP bound on charged scalar, i.e., $m_{H^+}>80\GeV$ \cite{Abbiendi:2013hk}. And also to satisfy the constraints from electroweak precision tests (EWPT) \cite{Machado:2015sha}, we further assume that the masses of neutrinophilic doublet scalars are degenerate as $m_{H^+}\!=\!m_{H_3}\!=\!m_{A}\!=\!m_{\Phi_\nu}$.

Since the heavy Majorana neutrino $N_R$ also exists in canonical
type-I seesaw \cite{type1}, searches for $N_R$ are already well
studied \cite{Keung:1983uu,Han:2006ip,delAguila:2008cj,Atre:2009rg,
Dev:2013wba,Das:2014jxa,Alva:2014gxa,Das:2012ze,Banerjee:2015gca,Das:2016hof}.
For more detail, see the recent review of neutrino and collider in
Ref. \cite{Deppisch:2015qwa} and references therein. Direct searches
for $N_R$ at colliders have also been performed at LEP
\cite{Abreu:1996pa,Achard:2001qv} and LHC
\cite{Chatrchyan:2012fla,Aad:2015xaa,Khachatryan:2016olu}. For
$m_N<m_W$, LEP has excluded the mixing $V_{\ell N}$ between the
heavy Majorana neutrino $N$ and the neutrino of flavor $\nu_\ell$
with $|V_{\ell N}|^2\gtrsim2\times10^{-5}$ \cite{Abreu:1996pa}. For
a more heavier $N_R$, LHC would give the most restrictive direct
limits. For instance, at $m_N=200\GeV$ the limit is $|V_{\ell
N}|^2<0.017$ and at $m_{N}=500\GeV$ the limit is $|V_{\ell
N}|^2<0.71$ \cite{Khachatryan:2016olu}. In $\nu$2HDM, the mixing
$V_{\ell N}$ is predicted as \cite{Perez:2009mu}:
\begin{equation}
V_{\ell N}=U_{\mbox{PMNS}} \hat{m}_\nu^{1/2} R~ m_N^{-1/2}\sim10^{-6}
\end{equation}
for EW-scale $m_N$, which is far below current limits.

\section{Phenomenology}

As pointed in introduction, the LNV scale is still unknown, hence
both low scale and high scale scenarios are allowed. For low scale
scenario, our model is a natural TeV-model, so it can be test at LHC
as we will discuss in SEC. \ref{HID} and \ref{coll}. On the other
hand, if the Majoron are assumed to be a DM candidate, a high LNV
scale $v_1\gtrsim10^4~\TeV$ are needed to satisfy constraints from
WMAP. In high scale scenario, the natural way to get correct
neutrino mass is keeping new scalar masses (i.e. $m_{H^+}$,
$m_{H_3}$ and $m_{A}$) around $v_1$ scale. So new scalars in case of
massive Majoron are out reach of LHC. The possible signatures of
massive Majoron will be discussed in SEC. \ref{MDM}.

\subsection{Invisible Higgs Decay}\label{HID}

Due to the existence of massless Majoron $J$, the Higgs scalar can
decay into Majorons though $H_{a}\to JJ$ and $H_{a}\to H_{b}
H_{b}\to 4J$ \cite{Bonilla:2015uwa,Bonilla:2015jdf,
Joshipura:1992ua,Joshipura:1992hp}. The Higgs-Majoron couplings are
derived as:
\begin{equation}
g_{H_a JJ}=\left(\frac{(O^I_{11})^2}{v_1}O^R_{1a}
+\frac{(O^I_{12})^2}{v_2}O^R_{2a}+\frac{(O^I_{13})^2}{v_3}O^R_{3a}\right)m_{H_a}^2,
\end{equation}
where $O^R$ and $O^I$ are the mixing matrices for CP-even and CP-odd scalars in Eq. \ref{or} and \ref{oi}. The partial decay width of $H_a\to JJ$ is then given by:
\begin{equation}
\Gamma(H_a\to JJ)=\frac{1}{32\pi}\frac{g_{H_a JJ}^2}{m_{H_a}}.
\end{equation}

Similar to the type-II seesaw case \cite{Bonilla:2015jdf}, the
smallness of $v_3$ indicates that the neutrinophilic doublet
$\Phi_\nu$ is basically decoupled. So we concentrate on the
invisible decays of $H_1$ and $H_2$.  Note that for light
$m_{H_3/A}<m_{H_2}/2$, $H_3$ or $A$ could also mediate a sizable
invisible decay of $H_2$\cite{Seto:2015rma}. The trilinear coupling
$H_2H_1H_1$ (see Eq. \ref{g211} in the appendix) contributes to
invisible decay of SM Higgs $H_2$ with $H_1\to JJ$. When
$m_{H_1}<m_{H_2}/2$, the decay mode $H_2\to H_1H_1$ would be
kinematically open. The partial decay width for $H_2\to H_1H_1$ is
computed as:
\begin{equation}
\Gamma(H_2\to H_1H_1)=\frac{g_{H_2H_1H_1}^2}{32\pi m_{H_2}}\left(1-\frac{4m_{H_1}^2}{m_{H_2}^2}\right)^{1/2}.
\end{equation}

For the decays of $H_1$ into SM particles, we refer Ref.
\cite{Robens:2015gla} for a more detail description. In this paper,
the benchmark points discussed in Sec. \ref{coco} are used to
illustrate the invisible decays of $H_1$ and $H_2$.
Varying the parameters, we find that the most relevant parameters
for the invisible Higgs decays are $\sin\alpha_{12}$, $v_1$, and
$m_{H_1}$. For simplicity, we fix the value of the following
parameters as:
\begin{eqnarray}\nonumber
m_{H^+}=m_{H_3}=m_{A}=m_{\Phi_\nu}=500~\GeV,\\
v_3=2~\MeV, \sin\alpha_{13}=\sin\alpha_{23}=2\times10^{-6},
\end{eqnarray}
meanwhile we vary the values of parameters $\sin\alpha_{12}$, $v_1$,
and $m_{H_1}$ as:
\begin{eqnarray}\nonumber
\sin\alpha_{12}&\in&[0,0.3],~v_1\in[500,1500],\\
m_{H_1}&\in&[10,100]~\GeV,~\text{for the low mass region},\\\nonumber
       &\in&[200,1000]~\GeV,~ \text{for the high mass region}.
\end{eqnarray}
It is checked that the above region is allowed by the boundedness
conditions in Eq. \ref{bd}. A fully scanning the whole parameter
space as done in Refs. \cite{Bonilla:2015uwa,Bonilla:2015jdf} is
worthwhile but beyond the scope of this paper. In the following, we
will give some qualitative discussion which is helpful to better
understand the scanning results of Refs.
\cite{Bonilla:2015uwa,Bonilla:2015jdf}.

{\textbf{1. High Mass Region:}}

\begin{figure}[!htbp]
\begin{center}
\includegraphics[width=0.33\linewidth]{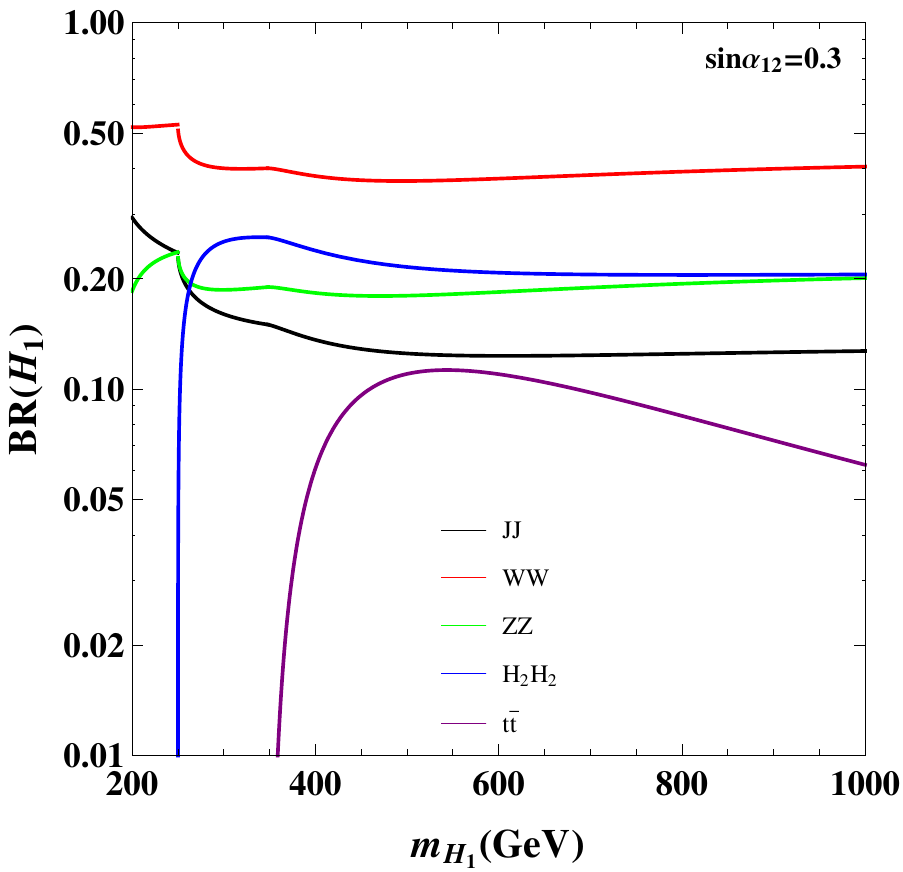}
\includegraphics[width=0.33\linewidth]{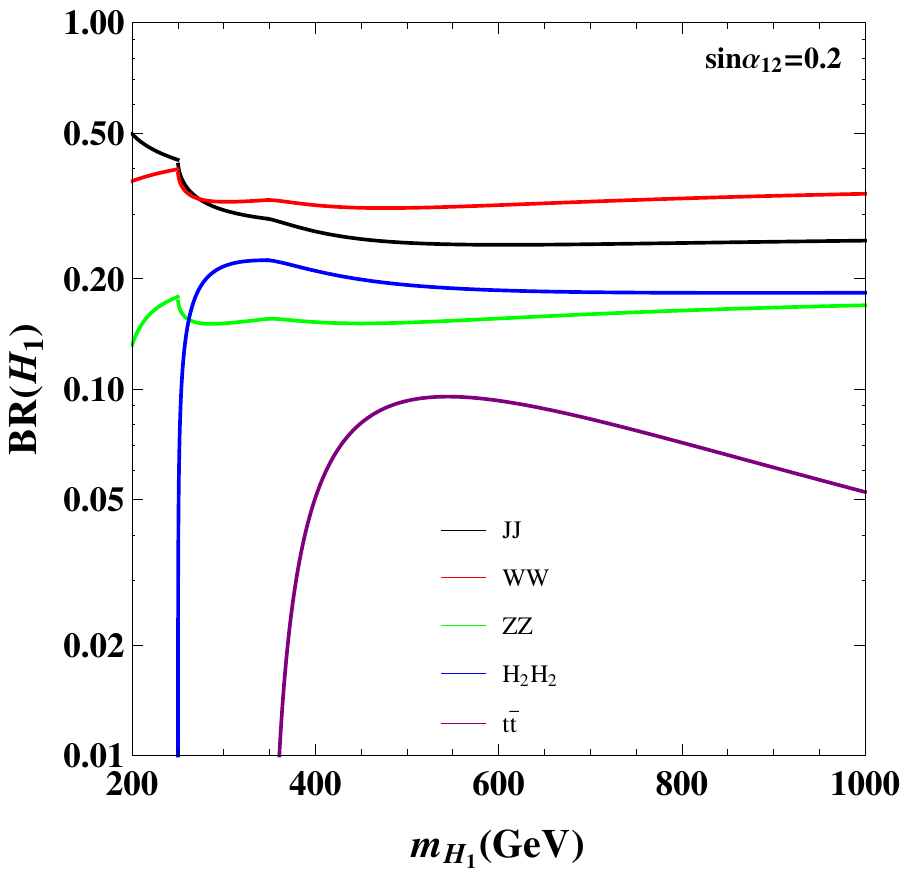}
\includegraphics[width=0.33\linewidth]{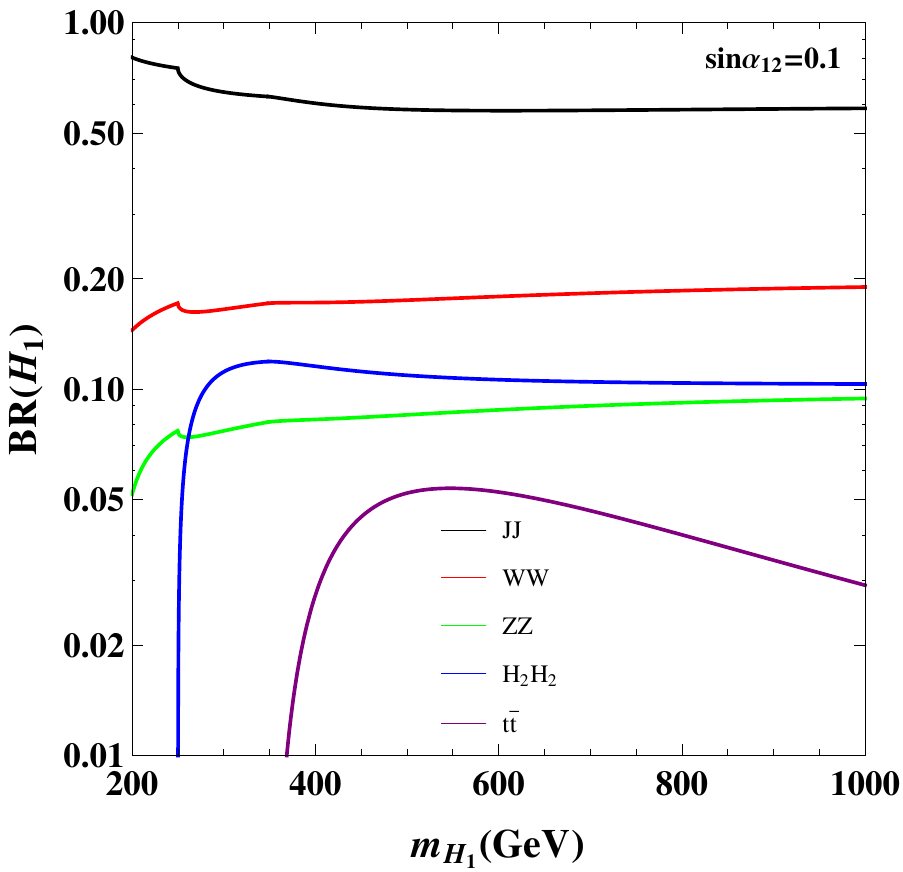}
\end{center}
\caption{Branching ratios of heavy singlet scalar $H_1$ as a function of $m_{H_1}$ for $\sin\alpha_{12}=0.3,0.2,0.1$, respectively. Here, we also set $v_1=1000\GeV$.
\label{BRH1-H}}
\end{figure}

First, we explore the high mass region $m_{H_1}>m_{H_2}$. In FIG. \ref{BRH1-H}, we show the branching ratios of $H_1$ as a function of $m_{H_1}$ for $\sin\alpha_{12}=0.3,0.2,0.1$ with $v_1=1000\GeV$. Clearly, the smaller $\sin\alpha_{12}$ is, the bigger BR($H_1\to JJ$) is, thus the bigger invisible decay of $H_1$ is. For $\sin\alpha_{12}=0.3$, BR($H_1\to JJ)\approx0.14$ when $m_{H_1}>350\GeV$, which is smaller than BR($H_1\to W^+W^-, H_2H_2, ZZ$). While for $\sin\alpha_{12}=0.1$, BR($H_1\to JJ)\approx0.6$, which is the dominant decay channel and makes $H_1$ quite different from the real scalar singlet in Refs. \cite{Robens:2015gla,Buttazzo:2015bka,Robens:2016xkb,Wang:2015saa}.  If $m_{H_1}>2 m_{H_2}$, then $H_1\to H_2H_2\to 4J$ will also contribute the invisible decay of $H_1$, but BR($H_1\to4J$) is expected to be less than $0.3\times0.23^2\approx0.016$. On the other hand, BR($H_1\to JJ$) is at least 0.14. Hence, the invisible decay of $H_1$ is dominant by $H_1\to JJ$.

\begin{figure}[!htbp]
\begin{center}
\includegraphics[width=0.32\linewidth]{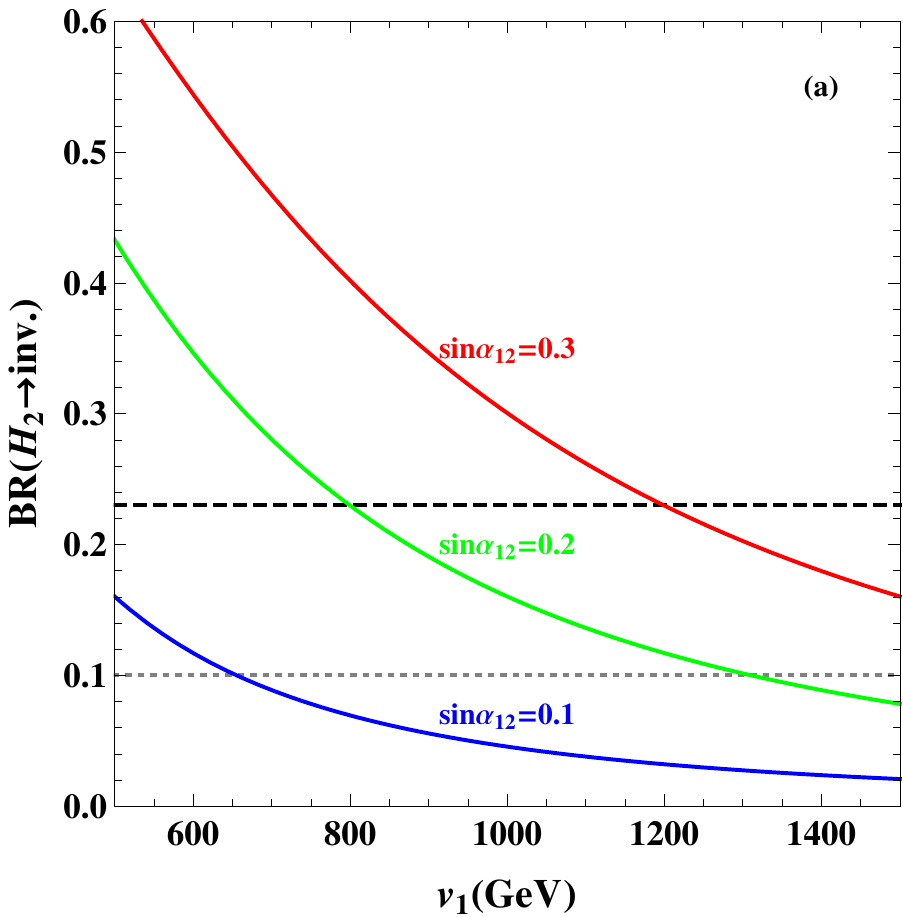}
\includegraphics[width=0.33\linewidth]{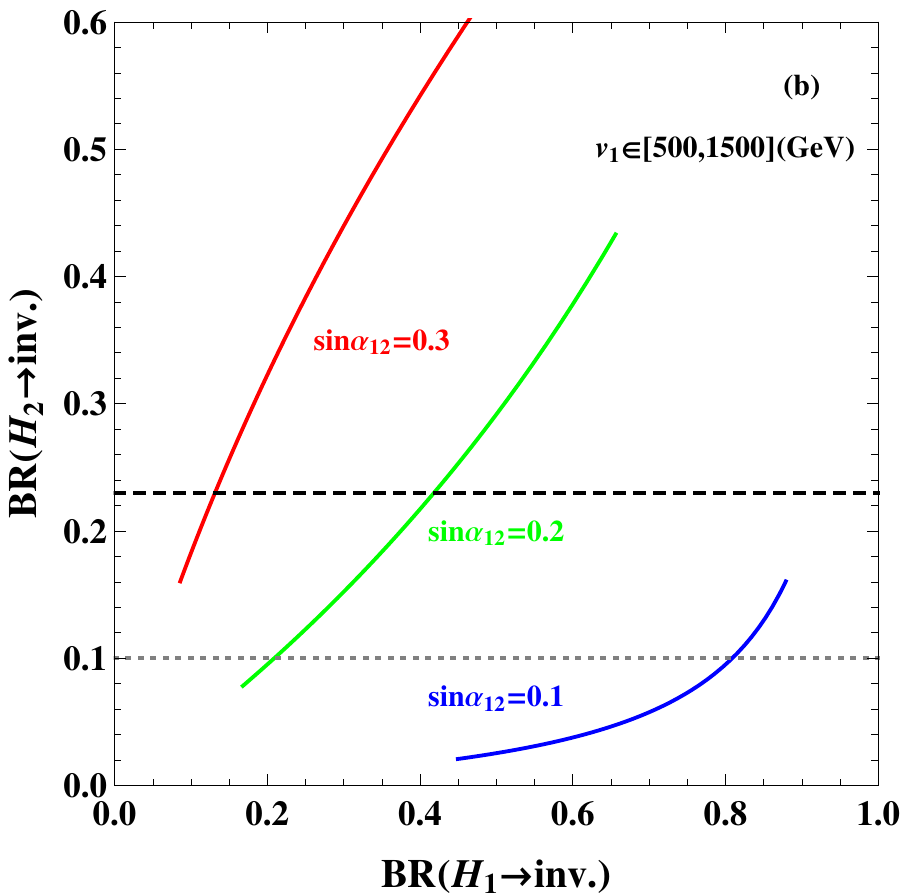}
\includegraphics[width=0.33\linewidth]{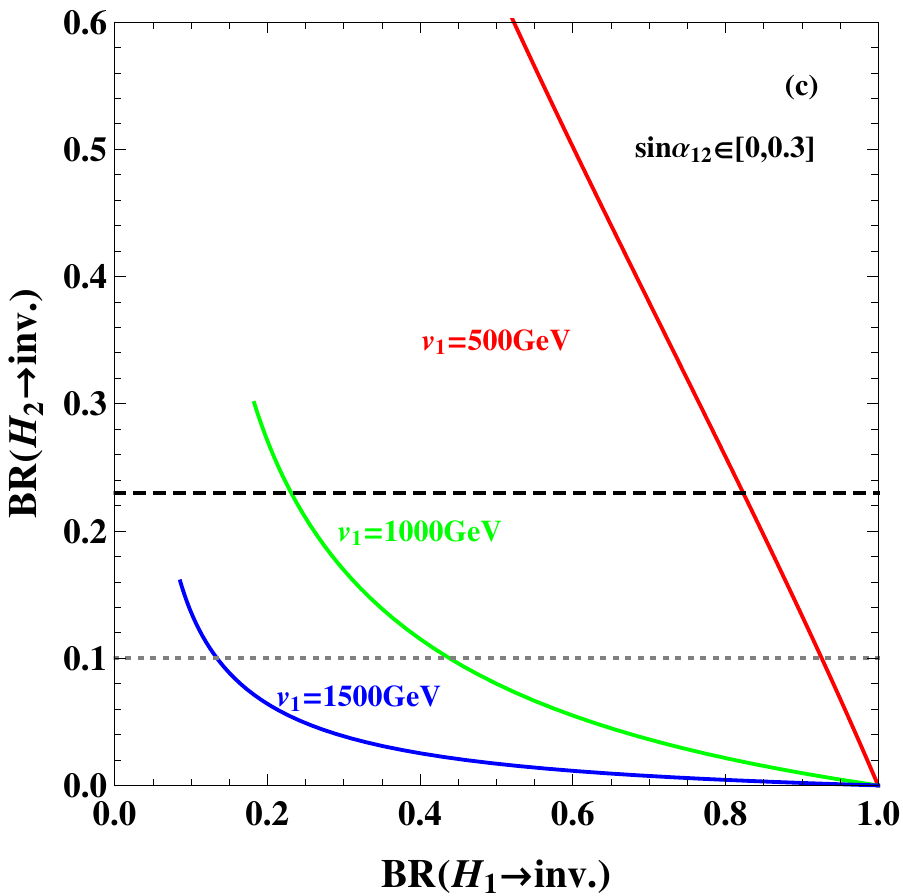}
\end{center}
\caption{(a) Branching ratios of invisible $H_2$ decay as a function of $v_1$ for $\sin\alpha_{12}=0.3,0.2,0.1$. (b) Relations between BR($H_1\to \mbox{inv.}$) and BR($H_2 \to \mbox{inv.}$) for $\sin\alpha_{12}=0.3,0.2,0.1$ by varying $v_1$ in the range of $500\!-\!1500\GeV$. (c) Relations between BR($H_1\to \mbox{inv.}$) and BR($H_2 \to \mbox{inv.}$) for $v_1=500,1000,1500\GeV$ by varying $\sin\alpha$ in the range of $0\!-\!0.3$.
In all these figures, we have set $m_{H_1}\!=\!300\GeV$. The dashed and dotted line correspond to current (0.23) \cite{Khachatryan:2014jba}  and future limit ($\approx0.1$) \cite{Okawa:2013hda} on BR($H_2 \to \mbox{inv.}$).
\label{H2inv-H}}
\end{figure}

In the high mass region with $m_{H_1}>m_{H_2}$, the invisible $H_2$ decay is dominant by $H_2\to JJ$. In FIG. \ref{H2inv-H} (a), we show BR($H_2\to \mbox{inv.}$)\footnote{inv. is short for invisible.} vs. $v_1$ for $\sin\alpha_{12}=0.3,0.2,0.1$. It is obvious that a smaller $\sin\alpha_{12}$ ($v_1$) will lead to smaller (larger) BR($H_2\to \mbox{inv.}$) for fixed $v_1$ ($\sin\alpha_{12}$). Considering current bound on BR($H_2\to \mbox{inv.}$)\cite{Khachatryan:2014jba}, $v_1\gtrsim1200(800)\GeV$ is required for $\sin\alpha_{12}=0.3(0.2)$. We then show the relations between BR($H_1\to \mbox{inv.}$) and BR($H_2 \to \mbox{inv.}$) with $m_{H_1}=300\GeV$ for $\sin\alpha_{12}=0.3,0.2,0.1$ by varying $v_1$ in the range of $500\!-\!1500\GeV$ in FIG. \ref{H2inv-H} (b), and for $v_1=500,1000,1500\GeV$ by varying $\sin\alpha$ in the range of $0\!-\!0.3$ in FIG. \ref{H2inv-H} (c). From these figures, we expect a positive(negative) correlation between BR($H_2\to \mbox{inv.}$) and BR($H_1\to \mbox{inv.}$) for varying $v_1$($\sin\alpha_{12}$).

{\textbf{2. Low Mass Region:}}

\begin{figure}[!htbp]
\begin{center}
\includegraphics[width=0.33\linewidth]{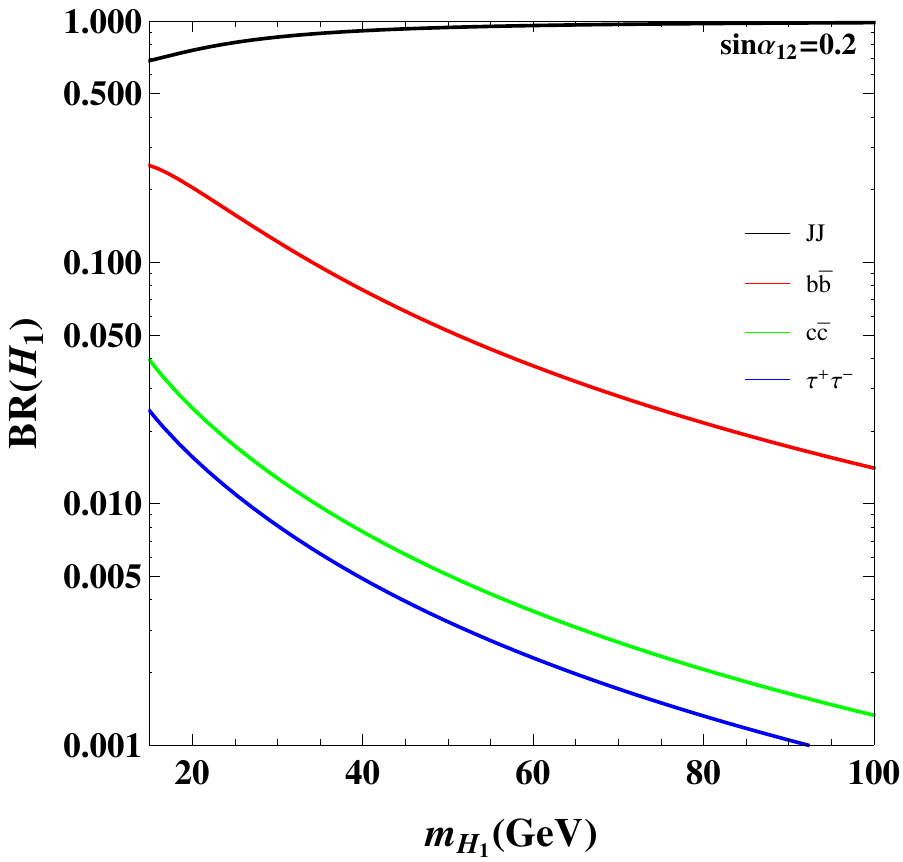}
\includegraphics[width=0.33\linewidth]{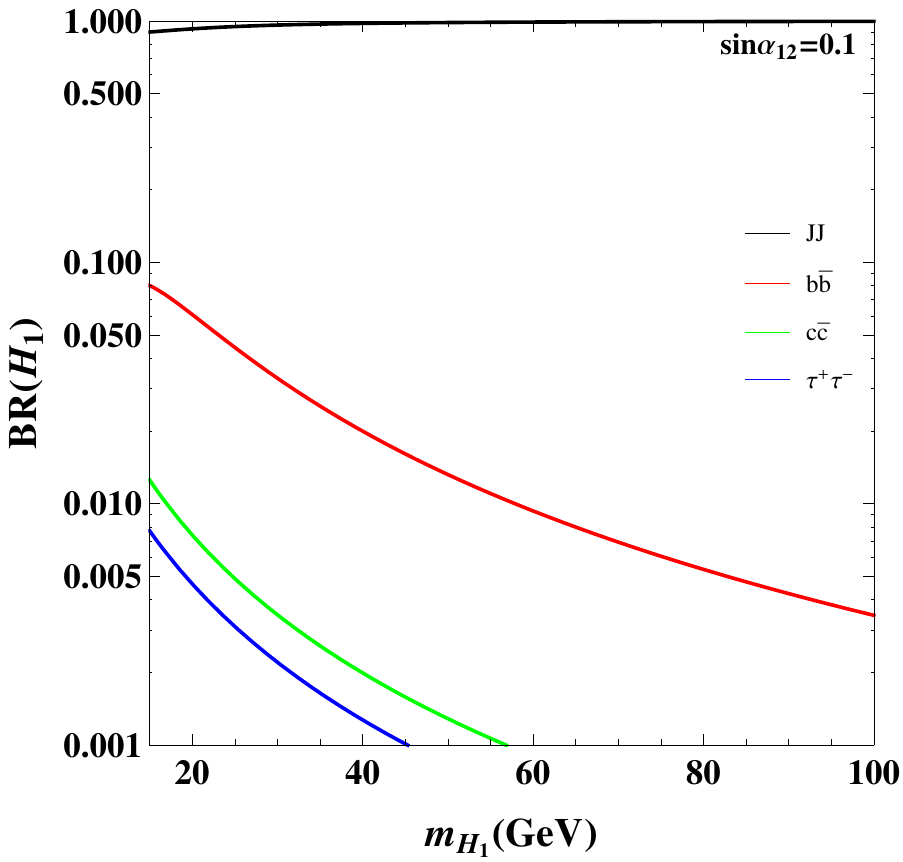}
\includegraphics[width=0.33\linewidth]{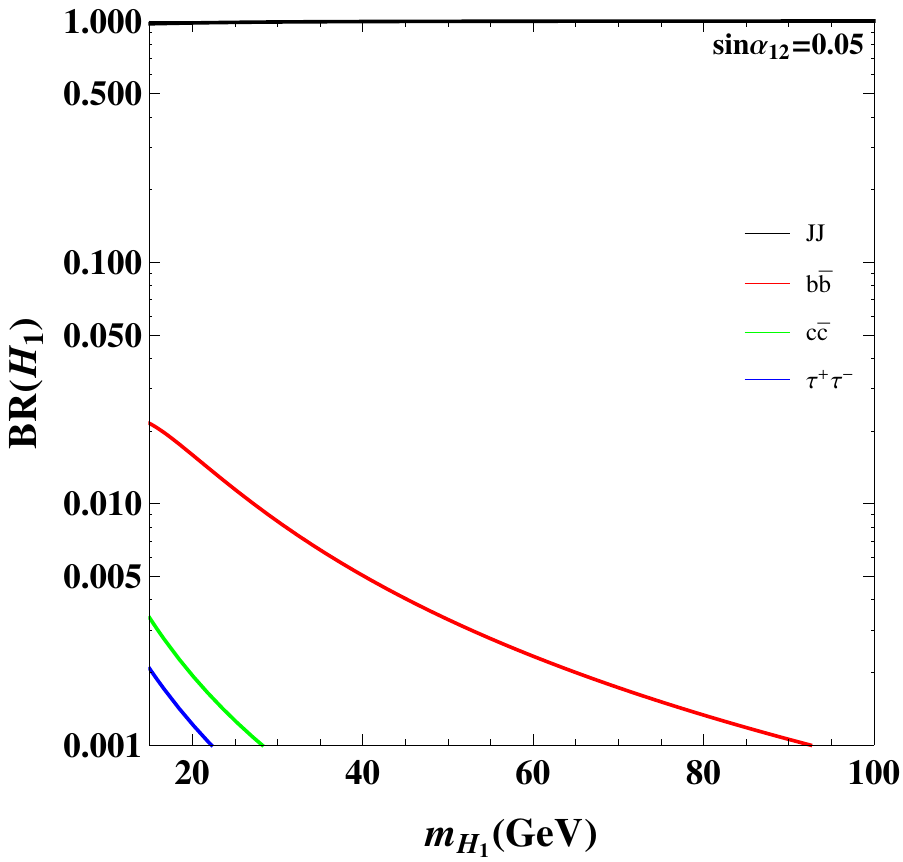}
\end{center}
\caption{Same as FIG. \ref{BRH1-H} but for $\sin\alpha_{12}=0.2,0.1,0.05$ in the low mass region of $m_{H_1}$.
\label{BRH1-L}}
\end{figure}

Then, we study the low mass region $m_{H_1}<m_{H_2}$. In FIG. \ref{BRH1-L}, we depict the branching ratios of $H_1$ as a function of $m_{H_1}$ for $\sin\alpha_{12}=0.3,0.2,0.1$ with $v_1=1000\GeV$.
Different from the high mass region, $H_1\to JJ$ is totally dominant in the low mass region for all allowed values of $\sin\alpha_{12}$. We expect BR($H_1\to \mbox{inv.})\gtrsim0.9$. The dominant visible decay of $H_1$ is $H_1\to b\bar{b}$, and BR($H_1\to b\bar{b}$) is typically  less than $0.1$. So exotic $H_2$ decays as $H_2\to H_1H_1\to b\bar{b}+\cancel{E}_T$ as well as $H_2\to H_1H_1\to 4b$ would be challenging at LHC \cite{Curtin:2013fra} for these benchmark points.

\begin{figure}[!htbp]
\begin{center}
\includegraphics[width=0.32\linewidth]{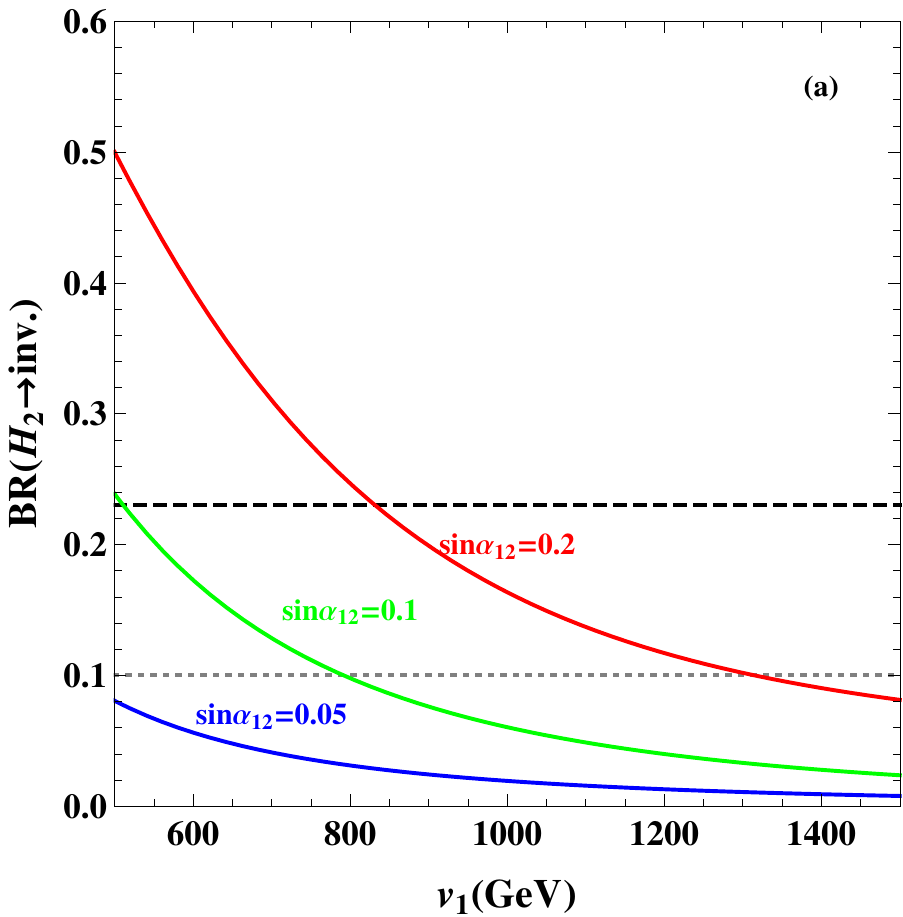}
\includegraphics[width=0.33\linewidth]{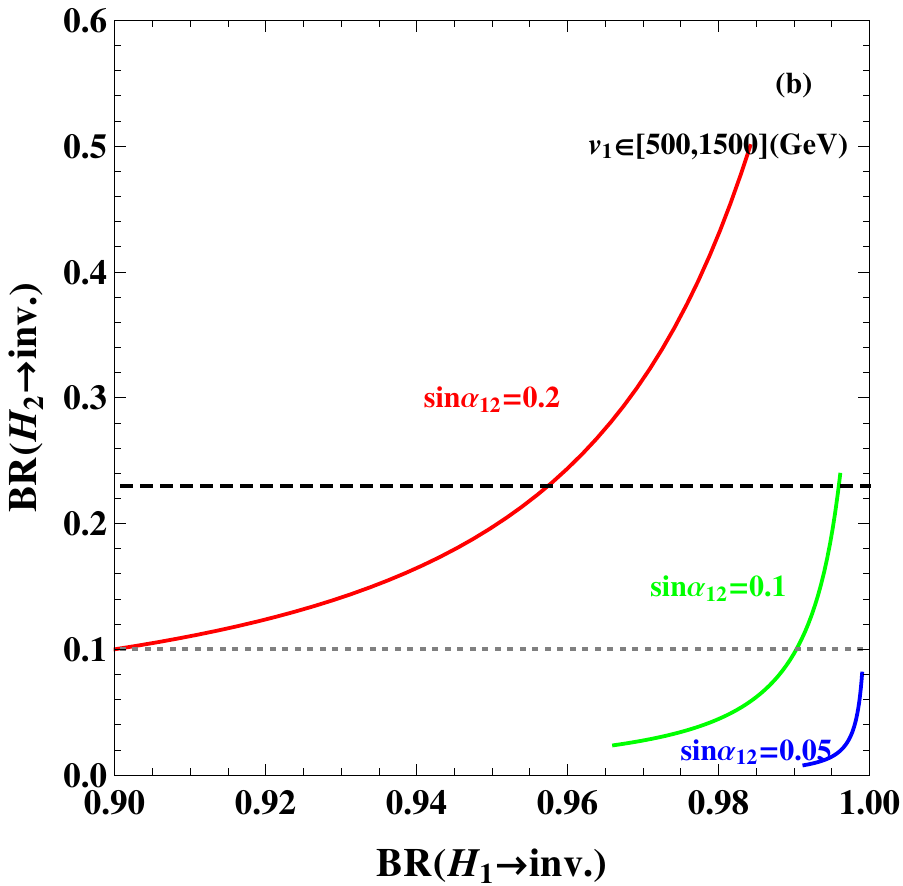}
\includegraphics[width=0.33\linewidth]{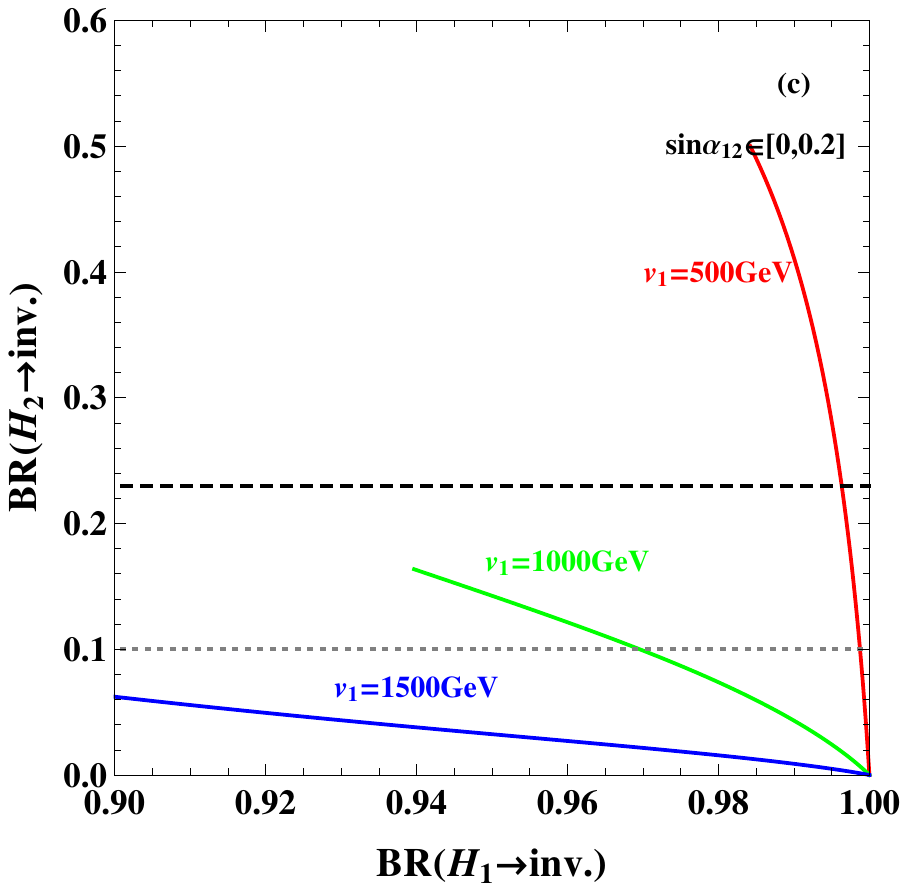}
\end{center}
\caption{Same as FIG. \ref{H2inv-H} but for $\sin\alpha_{12}=0.2,0.1,0.05$ with $m_{H_1}=50\GeV$.
\label{H2inv-L}}
\end{figure}

\begin{figure}[!htbp]
\begin{center}
\includegraphics[width=0.45\linewidth]{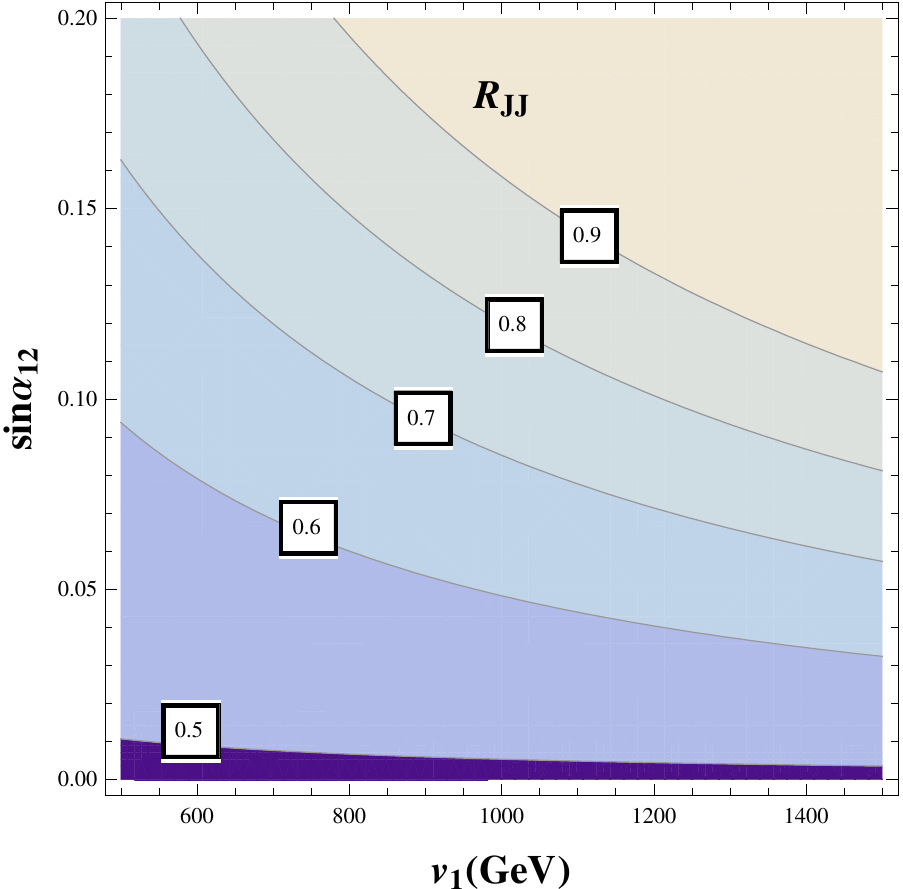}
\end{center}
\caption{Fraction of $H_2\to JJ$ to total invisible $H_2$ decay with $m_{H_1}=50\GeV$ in the $\sin\alpha_{12}$ vs. $v_1$ plane.
\label{RJJ}}
\end{figure}

In the low mass region with $m_{H_1}<m_{H_2}/2$, $H_2\to H_1H_1\to 4J$ will contribute to $H_2\to \mbox{inv.}$. Fixed $m_{H_1}=50\GeV$, we present BR($H_2\to \mbox{inv.}$) vs. $v_1$ in FIG. \ref{H2inv-L} (a), BR($H_2\to \mbox{inv.}$) vs. BR($H_1\to \mbox{inv.}$) for $\sin\alpha_{12}=0.2,0.1,0.05$ in FIG. \ref{H2inv-L} (b), and BR($H_2\to \mbox{inv.}$) vs. BR($H_1\to \mbox{inv.}$) for $v_1=500,1000,1500\GeV$ in FIG. \ref{H2inv-L} (c). All the qualitative arguments in the high mass region are also applicable here. But due to contributions of $H_2\to 4J$, bound on $v_1$ is slightly higher than it in the high mass case with same $\sin\alpha_{12}$.
Since both $H_2\to JJ$ and $H_2\to 4J$ contribute to $H_2\to \mbox{inv.}$, we quantize the contribution of $H_2\to JJ$ to $H_2\to \mbox{inv.}$ by defining:
\begin{equation}
R_{JJ} = \frac{\Gamma(H_2\to JJ)}{\Gamma(H_2\to \mbox{inv.})}
       = \frac{\Gamma(H_2\to JJ)}{\Gamma(H_2\to JJ)+\Gamma(H_2\to 4J)}.
\end{equation}
In FIG. \ref{RJJ}, we plot the contour of $R_{JJ}$ in the $\sin\alpha_{12}$ vs. $v_1$ plane. For $\sin\alpha_{12}>0.01$, $R_{JJ}>0.5$, which indicates that $H_2\to JJ$ is the dominant contribution to invisible $H_2$ decays in quite a large parameter space. And from FIG. \ref{RJJ}, we could conclude that the larger $v_1$ or the smaller $\sin\alpha$ is, the smaller the contribution of $H_2\to JJ$ to $H_2\to \mbox{inv.}$ is.

\subsection{Collider Signatures} \label{coll}

\begin{figure}[!htbp]
\begin{center}
\includegraphics[width=0.45\linewidth]{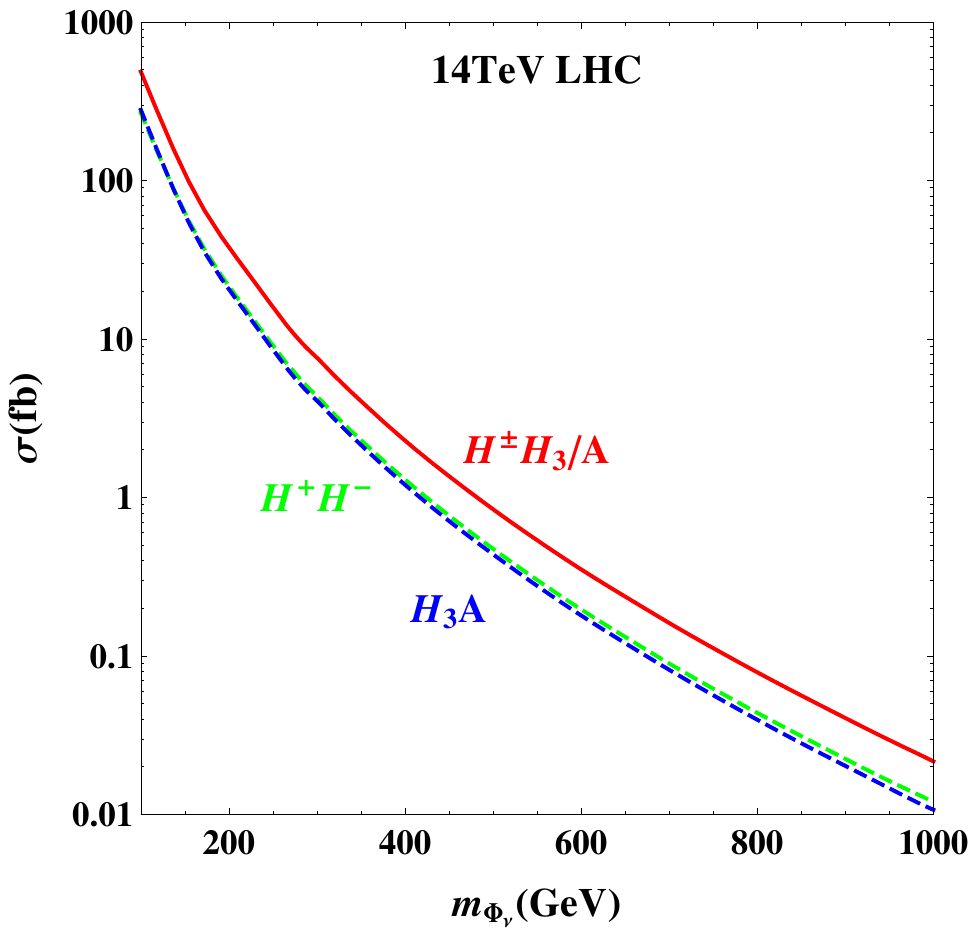}
\end{center}
\caption{Cross section of pair and associate production of the neutrophilic doublet scalars at 14 TeV LHC. We assume masses of the neutrophilic doublet scalars are degenerate as $m_{H^+}\!=\!m_{H_3}\!=\!m_A\!=\!m_{\Phi_\nu}$.
\label{CS}}
\end{figure}

Early papers on collider phenomenon of the $\nu$2HDM with heavy
Majorana neutrino $N_{Ri}$ can be found in Refs.
\cite{Ma:2000cc,Haba:2011nb,BarShalom:2008gt}, and they mainly
concentrate on the charged scalar $H^+$. Following Ref.
\cite{Haba:2011nb}, we give a brief discussion of the signatures at
LHC by taking into account the contribution of neutral scalars $H_3$
and $A$ in neutrinophilic 2HDM. In FIG. \ref{CS}, we show the cross
section of pair and associate production of the neutrinophilic
doublet scalars at 14 TeV LHC. Typically for EW-scale
$m_{\Phi_\nu}$, the cross sections are at the order of
$\mathcal{O}(\fb)$. The cross section of associate production $H^\pm
H_3/A$ is about twice larger than it of the pair production $H^+H^-$
or $H_3A$.

The decay properties of the neutrophilic doublet and the heavy Majorana neutrino are discussed in Ref. \cite{Haba:2011nb}. For $m_{\Phi_\nu}<m_N$, the dominant decay mode of $H^+$ is $H^+\to \ell^+\nu_i$ with $v_3\lesssim\mathcal{O}(\MeV)$ due to the mixing between light and heavy neutrino.
In this case, the most promising signatures at LHC is $H^+H^-\to \ell^+\ell^-+\cancel{E}_T$ \cite{Davidson:2010sf}. On the other hand for $m_{\Phi_\nu}>m_N$, the dominant decay mode of $H^+$ is $H^+\to \ell^+N_{Ri}$ with $v_3\lesssim\mathcal{O}(\GeV)$ \cite{Haba:2011nb}. The dominant decay mode of neutral scalars are reasonable to be $H_3\to \nu_{\ell} N_{Ri}$ and $A \to \nu_\ell N_{Ri}$, since they have the same Yukawa coupling as $H^+$. The heavy Majorana neutrino $N_{Ri}$ then decays as $N_{Ri}\to \ell^{\pm}W^{\mp}$, $N_{Ri}\to \nu_\ell Z$, $N_{Ri}\to \nu_\ell H_2$ \cite{Haba:2011nb,Atre:2009rg,Perez:2009mu,Basso:2008iv} \footnote{For simplicity, we set all the mixing angles among scalars to be zero here.}. For $m_N=200\GeV$, we have BR($N_{Ri}\to \ell^{\pm}W^{\mp}$)=0.60, BR($N_{Ri}\to \nu_\ell Z$)=0.28, and BR($N_{Ri}\to \nu_\ell H_2$)=0.12.

\begin{table*}[hbt]
\begin{tabular}{|l|l|l|}
\hline
 ~~A~: $H^+H^-\to \ell^+\ell^- N_{Ri} N_{Rj}$ & ~~B~: $H^\pm H_3/A\to \ell^\pm \nu N_{Ri} N_{Rj}$ &
 ~~C~: $H_3A\to \nu\nu N_{Ri} N_{Rj}$ \\
 \hline
 A.1: $\ell^+\ell^- \ell^\pm W^\mp \ell^\pm W^\mp$(0.360) & B.1: $\ell^\pm\nu \ell^\pm W^\mp \ell^\pm W^\mp$(0.360) & C.1: $\nu\nu \ell^\pm W^\mp \ell^\pm W^\mp$(0.360)\\
 A.2: $\ell^+\ell^- \ell^\pm W^\mp \nu Z$(0.336) & B.2: $\ell^\pm\nu\ell^\pm W^\mp \nu Z$(0.336) & C.2: $\nu\nu \ell^\pm W^\mp \nu Z$(0.336)\\
 A.3: $\ell^+\ell^- \ell^\pm W^\mp \nu H_2$(0.144) & B.3: $\ell^\pm\nu\ell^\pm W^\mp \nu H_2$(0.144)  & C.3: $\nu\nu\ell^\pm W^\mp \nu H_2$(0.144) \\
 A.4: $\ell^+\ell^- \nu Z \nu Z$(0.079) & B.4: $\ell^\pm\nu\nu Z \nu Z$(0.079) & C.4: $\nu\nu\nu Z \nu Z$(0.079) \\
 A.5: $\ell^+\ell^- \nu Z \nu H_2$(0.067) & B.5: $\ell^\pm\nu\nu Z \nu H_2$(0.067) & C.5: $\nu\nu\nu Z \nu H_2$(0.067)\\
 A.6: $\ell^+\ell^- \nu H_2 \nu H_2$(0.014) & B.6: $\ell^\pm\nu\nu H_2 \nu H_2$(0.014) & C.6: $\nu\nu\nu H_2 \nu H_2$(0.014)\\
\hline
\end{tabular}
\caption{Signals from pair and associate production of neutrophilic doublet $\Phi_\nu$ with their branching ratios given in the parentheses. Here, we set $m_N=200\GeV$.}
\label{Signal}
\end{table*}

In this paper, we concentrate on the case of $m_{\Phi_\nu}>m_N$. In TABLE \ref{Signal}, we summarize all the possible signatures (in $W^\pm,Z,H_2$ level) and classify them into three collum according to the production mechanism of $\Phi_\nu$. With $W^\pm,Z,H_2$ further decaying, there are various possible signatures. Due to the existence of heavy Majorana neutrino $N_{Ri}$, we concentrate on LNV  processes. The most interesting and distinct one is the same sign tri-lepton (SST) signature arising from B.1 of TABLE \ref{Signal} :
\begin{eqnarray}\nonumber
\mbox{B.1}\to \ell^\pm \nu \ell^\pm jj \ell^\pm jj \to 3\ell^\pm 4j+\cancel{E}_T
\end{eqnarray}
To our knowledge, such SST signature with $\Delta L=3$ can only take place in this model, thus it could be used to distinguish this model from other seesaw models.
There are also several same sign di-lepton (SSD) signatures with $\Delta L=2$:
\begin{eqnarray}
\mbox{B.2} &\to& \ell^\pm\nu \ell^\pm jj \nu jj \to 2\ell^\pm 4j+\cancel{E}_T \\
\mbox{B.3} &\to& \ell^\pm\nu \ell^\pm jj \nu jj \to 2\ell^\pm 4j+\cancel{E}_T \\
\mbox{C.1} &\to& \nu\nu \ell^\pm jj \ell^\pm jj \to 2\ell^\pm 4j+\cancel{E}_T
\end{eqnarray}
All these three processes contribute to the SSD signature $2\ell^\pm
4j+\cancel{E}_T$. And there is also a four lepton signature with
$\Delta L=2$:
\begin{equation}
\mbox{A.1} \to \ell^+\ell^- \ell^\pm jj \ell^\pm jj \to 3\ell^\pm \ell^\mp 4j
\end{equation}

\begin{figure}[!htbp]
\begin{center}
\includegraphics[width=0.45\linewidth]{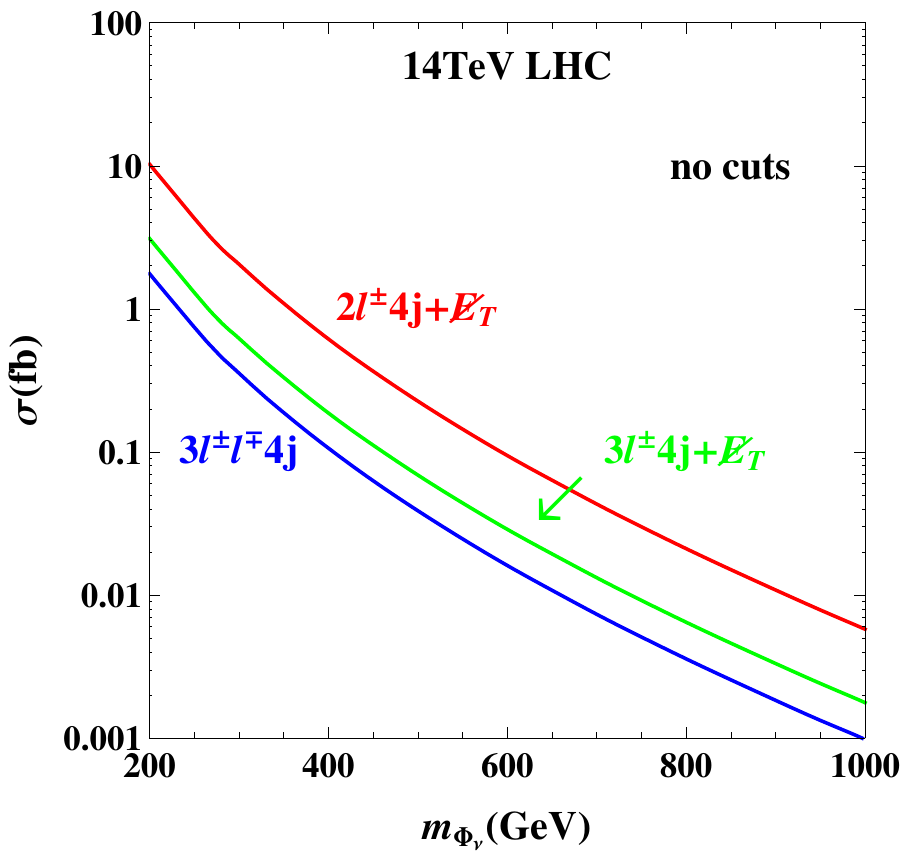}
\end{center}
\caption{Cross section of lepton number violation signatures at 14 TeV LHC. We also fix $m_N=200\GeV$.
\label{LNV}}
\end{figure}

In FIG. \ref{LNV}, we shows the theoretical cross section for the LNV signatures at 14 TeV LHC.
The SSD signature $2\ell^\pm4j+\cancel{E}_T$ has the largest cross section, but it also suffers a relative large background from $t\bar{t}W$. On the contrary, the four lepton signature $3\ell^\pm\ell^\mp4j$ has a relative clean background, but its cross section is the smallest. The SST signature $3\ell^\pm 4j+\cancel{E}_T$ seems very promising, since it is nearly background free. Thus it might be testable for $m_{\Phi_\nu}\lesssim700\GeV$ with integrated luminosity of $300\fb^{-1}$ at 14 TeV LHC. A fully discussion and simulation of these LNV signatures at LHC will be carried out in another paper \cite{Han:2016ly}.

\subsection{Majoron Dark Matter} \label{MDM}

Considering the non-perturbative gravitational effects, the Majoron
$J$ could get an $\mathcal{O}(\keV)$ mass
\cite{Coleman:1988tj,Kallosh:1995hi}, and play the role of decaying
dark matter \cite{Lattanzi:2007ux,Bazzocchi:2008fh,Esteves:2010sh}.
It is possible to realize EW-scale decaying
\cite{Gu:2010ys,Queiroz:2014yna} or stable dark matter
\cite{Lindner:2011it,Chang:2014lxa,Chang:2016pya} in Majoron models.
In this paper, we focus on $\mathcal{O}(\keV)$ Majoron and
corresponding phenomenon.

\begin{figure}[!htbp]
\begin{center}
\includegraphics[width=0.6\linewidth]{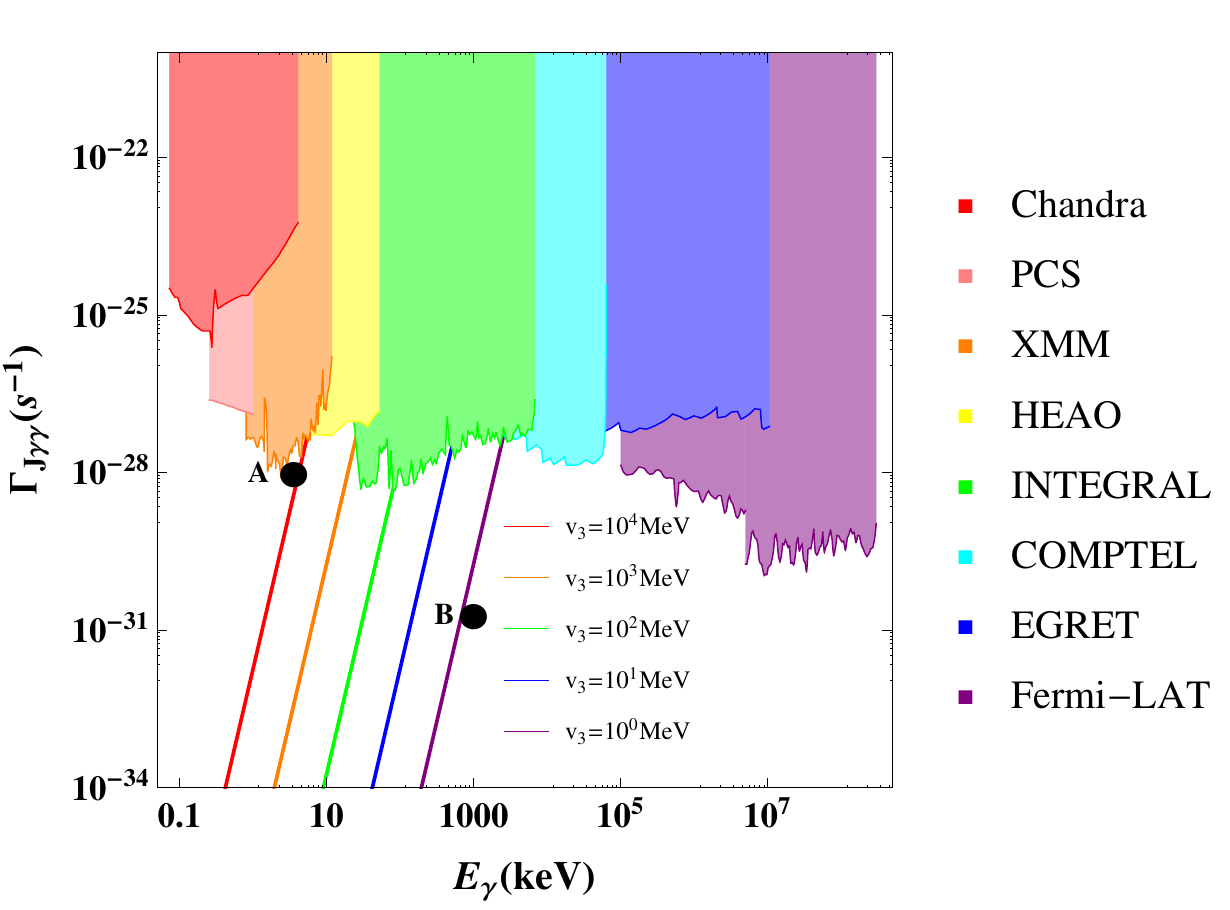}
\end{center}
\caption{The predicted decay rate of $J\to\gamma\gamma$ as a function of $E_\gamma$ for different values of $v_3$. Point A and B are the benchmark points to interpret the $3.5\keV$ and $511\keV$ line excess. The constraints are (from left to right): CHANDRA Low Energy Transmission Grating (LETG) observations of NGC3227 (red) \cite{Bazzocchi:2008fh}, the Milky Way halo observed with PCS (pink) \cite{Boyarsky:2006hr}, XMM observations of the Milky Way and M31 (orange) \cite{Boyarsky:2006ag}, the diffuse x-ray background observed with HEAO  (yellow) \cite{Boyarsky:2005us}, INTEGRAL  diffuse background (green)\cite{Boyarsky:2007ge}, COMPTEL search (cyan) \cite{Yuksel:2007dr}, EGRET search (blue) \cite{Strong:2004de}, Fermi-LAT $\gamma$-ray searches (purple) \cite{Ackermann:2012qk}.
\label{xraysg}}
\end{figure}

For decaying Majoron dark matter, the present majoron density can be expressed as:
\begin{equation}
\Omega h^2 = \beta \left(\frac{m_J}{1.25~\keV} \right)e^{-t_0/\tau_J},
\end{equation}
where $h$ is the Hubble constant, $t_0$ is the age of the universe, and $\beta$ is in the range $10^{-5}\!-\!1$ corresponding to the majoron thermal history \cite{Frigerio:2011in}. The decay mode of Majoron $J$ is dominant by $J\to \nu\nu$.
Induced by the $k$-term in Eq. \ref{vphi}, the Majoron $J$ has non-zero component along the SM and neutrinophilic doublet, and it is approximately given by:
\begin{equation}\label{Jpro}
J\sim I_1 + \frac{2 v_3^2}{v_1 v_2}I_2 - \frac{2 v_3}{v_1} I_3
\end{equation}
According to this, we can derive the Majoron-neutrino coupling (to leading order) \cite{Schechter:1981cv}:
\begin{equation}
g_{J\nu_i\nu_j}= -\frac{2 m_i^\nu}{v_1} \delta_{ij}+ ...
\end{equation}
and the corresponding decay width:
\begin{equation}\label{Jvv}
\Gamma_{J\to\nu\nu}= \frac{m_J}{2\pi} \frac{\sum_i (m_i^\nu)^2}{v_1^2}
\end{equation}
The late decay $J\to \nu\nu$ would produce too much power at large scales, thus spoiling the CMB anisotropy spectrum. WMAP third year data has set an upper limit \cite{Bazzocchi:2008fh,Lattanzi:2008ds}:
\begin{equation}\label{WMAP}
\Gamma_J < 6.4\times 10^{-19} s^{-1},~\mbox{with}~ 0.12\keV<\beta m_J<0.17\keV.
\end{equation}
From Eq. \ref{Jvv}, it is clear that such limit can be easily satisfied as long as $v_1$ is large enough. For instance, an $\mathcal{O}(\keV)$ Majoron requires $v_1\gtrsim\mathcal{O}(10^4\TeV)$ to satisfy the WMAP limit. In the following discussion, we take $v_1$ to saturate the upper limit on $J\to\nu\nu$. Since $\beta\in[10^{-5},1]$, then $m_J\sim0.1\!-\!10^4\keV$. More interesting, the sub-leading decay mode of $J$ is $J\to \gamma\gamma$, which is mediated by charged fermions at one-loop level:
\begin{equation}\label{JAA}
\Gamma_{J\to\gamma\gamma} = \frac{\alpha^2m_J^3}{64\pi^3}\left|\sum_f N_f Q_f^2
\frac{2 v_3^2}{v_2^2 v_1}(-2T_3^f)\frac{m_J^2}{12m_f^2}\right|^2,
\end{equation}
where $N_f$, $Q_f$, $T_3^f$ and $m_f$ are the color factor, electric charge, weak isospin and mass of SM fermion $f$, respectively. Note from Eq. \ref{JAA} that, $\Gamma_{J\to\gamma\gamma}$ only depends on $v_3$ with fixed values of $m_J$ and $v_1$. The predicted decay rate of $J\to\gamma\gamma$ as a function of $E_\gamma(=m_J/2)$ for different values of $v_3$ is shown in FIG. \ref{xraysg}. It is clear that a larger $v_3$ leads to a larger $\Gamma_{J\to\gamma\gamma}$. Since the performed line emission search has already excluded $\Gamma_{J\to\gamma\gamma}\gtrsim\mathcal{O}(10^{-28}s^{-1})$, a larger $v_3$ actually prefers a smaller $E_\gamma$ for the survived $\gamma$-rays. Further considering the LFV bound on $v_3\gtrsim\mathcal{O}(\MeV)$ with EW-scale $m_{\Phi_\nu}$, the predicted $E_\gamma$ is usually $\lesssim10\MeV$, which covers the right region of $m_J$ to satisfy the relic density of decaying dark matter \cite{Frigerio:2011in}.

As discussed in Sec. \ref{intro}, the Majoron DM is also a good candidate to explain several $\keV$-line excesses.  Here, we have chose two different benchmark points to interpret the observed $3.5~\keV$ and $511~\keV$ line excesses respectively.
First, the direct decay mode $J\to \gamma\gamma$ for $\keV$-scale Majoron can be used to interpret the $3.5~\keV$ line excess with \cite{Queiroz:2014yna,Bulbul:2014sua}:
\begin{equation}
m_J\sim7\keV,~\mbox{and}~\Gamma_{J\to\gamma\gamma}\sim10^{-28}s^{-1},
\end{equation}
which corresponds to benchmark point A in FIG. \ref{xraysg}. Such
requirement can be satisfied with $v_3\sim10\GeV$ and $v_1\sim
10^4\TeV$.  Note that $v_3$ in this range can satisfy
the tight astrophysical constraints for $v_1\sim10^4\TeV$, as well
as the direct x-ray search bounds in FIG. \ref{xraysg} and WMAP
limit in Eq. \ref{WMAP}.

Second, for $\MeV$-scale $m_J$, the decay mode $J\to e^+e^-$ is potential to  explain the $511~\keV$ line excess with the requirement \cite{Knodlseder:2003sv,Picciotto:2004rp,Hooper:2004qf}:
\begin{equation}\label{Jee1}
\Gamma_{J\to e^+e^-}^{\tiny\mbox{exp}}\simeq 6.3\frac{m_J}{1\MeV}\times10^{-27} s^{-1},
\end{equation}
where we have assume that the Majoron DM $J$ accounts for all the observed DM relic density. In our model, the decay width of $J\to e^+e^-$ is given by:
\begin{equation}\label{Jee2}
\Gamma_{J\to e^+e^-} = \frac{m_J}{8\pi}\left|\frac{2v_3^2}{v_1 v_2}\frac{m_e}{v_2}\right|^2
\left(1-4\frac{m_e^2}{m_J^2}\right)^{1/2}.
\end{equation}
Combine Eq. \ref{Jee1} and \ref{Jee2}, we have:
\begin{equation}
\frac{\Gamma_{J\to e^+e^-}}{\Gamma_{J\to e^+e^-}^{\tiny\mbox{exp}}} =
\left(\frac{v_3^2}{v_1v_2}\right)^2\left(1-4\frac{m_e^2}{m_J^2}\right)^{1/2}\times1.7\times10^{35}.
\end{equation}
 Taking $m_J=2\MeV$, the required decay width can be
obtained for $v_3\sim1\MeV$ and $v_1\sim10^6\TeV$. Meanwhile the
WMAP limit on $\Gamma_J$ in Eq. \ref{WMAP} can be satisfied and the
decay width of $\Gamma_{J\to\gamma\gamma}$ corresponding to
benchmark point B in FIG. \ref{xraysg} is far below current direct
x-ray limits. Note that to acquire $v_3\sim1\MeV$, we also need
$m_{\Phi_\nu}\gtrsim\TeV$ to satisfy LFV constraints.

In principle, the discussions for invisible Higgs decay and LHC
signatures in previous case for massless $J$ are still applicable
for Majoron DM, since $J$ is still invisible at LHC and much lighter
than electroweak scale. But with such large $v_1\gtrsim10^4~\TeV$ to
satisfy WMAP limit, the coupling of $H_a JJ$ is so small, thus the
branching ratio of invisible Higgs decay is tiny. On the other hand,
the masses of $\Phi_{\nu}$ gets a large contribution from
$\beta_3$-term in the scalar potential and would be much heavier
than TeV-scale, thus beyond the reach of LHC.

\section{Conclusion}\label{ccl}

In this paper, we propose a new model to realize the spontaneous
violation of global $U(1)_L$ symmetry in the context of $\nu$-2HDM,
where a neutrinophilic doublet scalar $\Phi_\nu$ with lepton number
$L=1$, a complex singlet scalar $\sigma$ with $L=1/2$, and neutral
right-handed fermion singlets $N_{Ri}$ with $L=0$ are introduced in
addition to SM particles. The global $U(1)_L$ symmetry is
spontaneously broken by the VEV of $\sigma$, which leads to an
(nearly) massless Majoron $J$ and also induces a small VEV of
$\Phi_\nu$. Neutrino masses are generate at tree level type-I seesaw
like diagram with the SM doublet $\Phi$ replaced by the
neutrinophilic doublet $\Phi_\nu$. Due to the smallness of
$\langle\Phi_\nu\rangle$, the model is naturally an
$\mathcal{O}(\TeV)$ scale seesaw, and thus detectable in the reach
of LHC.

Constraints coming from astrophysics, lepton flavor violation, and
direct collider searches are taking into account. The astrophysical
constraints set an upper limit on VEV of $\Phi_\nu$, i.e.,
$v_3\lesssim0.09\GeV$ for the LNV scale $v_1$ at $1\TeV$. On the
other hand, the LFV constraints set a lower limit on $v_3$, i.e.,
$v_3\gtrsim1\MeV$ for $m_{\Phi_\nu}=600\GeV$. Due to the existence
of heavy $N_R$ in Majorana case of $\nu$-2HDM, we explain the huge
enhancement ($\sim10^6$) of lower limit on $v_3$ comparing to Dirac
case of $\nu$-2HDM.  Based on various signals arising from new
particles in our model, we investigate the direct search limits
carried out by LEP and LHC as well. By choosing proper parameters,
we find  EW-scale new particles are allowed and some benchmark
points are given to illustrate the phenomenological feature of the
model.

For massless Majoron, two aspects of LHC signatures are studied: the
invisible Higgs decays and LNV signatures. The invisible Higgs
decays can be induced by $H_a\to JJ$ and $H_a\to H_bH_b\to4J$. In
the decoupling limit of $\Phi_\nu$, the two dominant variable that
have impact on invisible Higgs decays are $\sin\alpha_{12}$ and
$v_1$. Comparing several benchmark points, we conclude that the
Majoron could induce large invisible Higgs decay and future
experiments prefer smaller $\sin\alpha_{12}$ and larger $v_1$. The
$\nu$-2HDM with $N_{R}$ has three kinds of LNV signatures. The most
interesting and distinct one is the same sign tri-lepton signature
$3\ell^\pm4j+\cancel{E}_T$, which can be used to distinguish from
other seesaw models. The other two LNV signatures
$2\ell^\pm4j+\cancel{E}_T$ and $3\ell^\pm\ell^\mp4j$ are also
promising to test this model at LHC.

Finally, the Majoron with
$m_J\sim\mathcal{O}(\keV)-\mathcal{O}(\MeV)$ mass is considered. In
this case, the Majoron can serve as a good decaying dark matter
candidate. To fulfill the CMB constraints on $\Gamma_{J\to \nu\nu}$,
the LNV scale is required to be $\mathcal{O}(10^3-10^6\TeV)$. The
sub-leading decay mode $J\to \gamma\gamma$ is also calculated and
compared with current experiments. We find the current limits have
already excluded some parameter space.  Further, we
point out that $J\to \gamma\gamma$ with $m_J\sim7~\keV$ can explain
the $3.5~\keV$ line excess and $J\to e^+e^-$ with
$m_J\sim\mathcal{O}(\MeV)$ can interpret the $511~\keV$ line excess.
Two different benchmark points are given to illustrate these two
excesses respectively.

\section*{Acknowledgement}

The work of Weijian Wang is supported by National Natural Science Foundation
of China under Grant Numbers 11505062, Special Fund of Theoretical Physics under
Grant Numbers 11447117 and Fundamental Research Funds for the Central Universities.

\section*{Appendix}

The coupling of $H_2H_1H_1$:
\begin{eqnarray}\label{g211}\nonumber
\frac{g_{H_2H_1H_1}}{2} &=& 3\lambda_1 v_2 (O^R_{12})^2 O^R_{22}+3\lambda_1 v_3  (O^R_{13})^2 O^R_{23} + 3 \beta_1 v_1 (O^R_{11})^2 O^R_{21}  \\\nonumber
&+&\frac{\lambda_3+\lambda_4}{2}\left[(O^R_{13})^2O^R_{22}v_2+(O^R_{12})^2O^R_{23}v_3
+2 O^R_{12}O^R_{13}(O^R_{23}v_2+O^R_{22}v_3)\right] \\
&+& \frac{\beta_2}{2}\left[(O^R_{12})^2O^R_{21}v_1+(O^R_{11})^2O^R_{22}v_2 +
 2 O^R_{11}O^R_{12}(O^R_{22} v_1 + O^R_{21} v_2) \right] \\\nonumber
&+& \frac{\beta_3}{2}\left[(O^R_{13})^2O^R_{21}v_1+(O^R_{11})^2O^R_{23}v_3 +
 2 O^R_{11}O^R_{13}(O^R_{23} v_1 + O^R_{21} v_3) \right] \\\nonumber
&-&\frac{k}{2}\left[(O^R_{11})^2O^R_{22}v_3+(O^R_{11})^2O^R_{23}v_2+ 2 O^R_{11}O^R_{12}(O^R_{21}v_3 + O^R_{23} v_1) + 2 O^R_{11}O^R_{13}(O^R_{21}v_2+O^R_{22}v_1) \right].
\end{eqnarray}

\end{document}